\documentclass[aps,prc,twocolumn,superscriptaddress]{revtex4-2}
\usepackage{CJKutf8}
\usepackage{amssymb}
\usepackage{amsmath}
\usepackage{diagbox}
\usepackage{makecell}
\usepackage{tikz}
\usepackage{color}
\usepackage{graphicx} 
\usepackage{multirow} 
\usepackage{makecell}
\usepackage[hidelinks]{hyperref}
\begin{document}

\title{Dipole response of deformed halo nuclei $^{31}$Ne and $^{37}$Mg}

\author{Xiao Lu } 
\email[]{luxiao@itp.ac.cn}
\affiliation{CAS Key Laboratory of Theoretical Physics, Institute of Theoretical Physics, Chinese Academy of Sciences, Beijing 100190, China} 
\author{Hiroyuki Sagawa } 
\email[]{hiroyuki.sagawa@gmail.com}
\affiliation{CAS Key Laboratory of Theoretical Physics, Institute of Theoretical Physics, Chinese Academy of Sciences, Beijing 100190, China}
\affiliation{RIKEN Nishina Center for Accelerator-Based Science, Wako 351-0198, Japan} 
\affiliation{Center for Mathematics and Physics University of Aizu, Aizu Wakamatsu, Fukushima 965-0001, Japan} 
\author{Shan-Gui Zhou } 
\email[]{sgzhou@itp.ac.cn}
\affiliation{CAS Key Laboratory of Theoretical Physics, Institute of Theoretical Physics, Chinese Academy of Sciences, Beijing 100190, China} 
\affiliation{School of Physical Sciences, University of Chinese Academy of Sciences, Beijing 100049, China} 
\affiliation{School of Nuclear Science and Technology, University of Chinese Academy of Sciences, Beijing 100049, China}  

\date{\today}

\begin{abstract} 
We study the soft electric dipole ($E1$) response of deformed halo nuclei $^{31}$Ne and $^{37}$Mg using a deformed Woods-Saxon potential, with the potential depth adjusted to reproduce empirical separation energy of last neutron orbit, i.e., 150 keV for $^{31}$Ne and 220 keV for $^{37}$Mg. 
The configuration dependence of the $E1$ strength near the neutron  threshold is pointed out.
The halo configurations $[321]3/2$ at $\beta_2=0.5$ and $[330]1/2$ at $\beta_2=0.24$ in $^{31}$Ne contain large amplitudes of halo $p$-shell orbits,  which significantly enhance the threshold strength by several times compared to the non-halo configuration $[202]5/2$ at $\beta_2=0.32$. 
In $^{37}$Mg, the last neutron configuration is assigned as $[321]1/2$ at a large deformation of $\beta_2=0.46$, which involves a halo $p$-shell configuration that  significantly enhances the soft dipole strength. This enhancement is about 60\%   larger than that of the $[321]3/2$ configuration in $^{31}$Ne because of large $p$-shell probability in $^{37}$Mg. 
 Experimental confirmation of the soft dipole strength is highly desired
to determine the deformation and the configuration of the
last neutron orbits both  in $^{31}$Ne  and $^{37}$Mg.
\end{abstract}


\maketitle
\section{Introduction}
Structural evolution of atomic nuclei towards the neutron and proton drip lines in the nuclear chart is one of the most important and intriguing issues in current nuclear physics.  In particular,
the location of the neutron drip line is the key to understanding the stability of  the  nuclear many-body quantum system.  Nuclei exhibit a peculiar
feature of nuclear structure, such as the neutron halo nuclei found in the vicinity of the neutron drip line \cite{TANIHATA2013215,Zhou:2017ldu,ye24}.
A halo structure appears as an extended density distribution, which is one of the characteristic features of weakly bound
neutron-rich and proton-rich nuclei. It was first observed in $^{11}$Li, which exhibits an abnormally large interaction cross section \cite{Tanihata1985}.
Other examples include $^{11}$Be \cite{TANIHATA1988,FUKUDA1991} and $^{19}$C \cite{19c1995,Nakamura1999}, both
of which are typical one-neutron ($1n$) halo nuclei. 
Recently, the $1n$ halo nuclei $^{31}$Ne \cite{Nakamura2009,Nakamura2014} and $^{37}$Mg \cite{mg37,mg371}, as well as the $2n$ halo nuclei $^{17}$B \cite{17B02,17B21}, $^{19}$B \cite{19B20}, $^{22}$C \cite{Tanaka2010}, and $^{29}$F \cite{29F20} have been identified. 
A key question is how these heavier halo nuclei can be formed and in which regions of the nuclear chart they occur, particularly when discussing nuclear structure near the drip line.


The halo is formed under the condition that the nucleus has a very small $1n$ or $2n$ separation energy $(S_n, S_{2n}\leq 1$ MeV) and that the least bound neutron has low  angular momentum  $l\leq 1$ \cite{SAGAWA1992,RIISAGER1992,meng96,MENG06,TANIHATA2013215,Meng_2015}. The latter condition is not satisfied  for nuclei with $20 < N < 28$ when adopting the conventional single-particle shell model. This is because the least bound neutron in the $1f_{7/2}$ orbit is affected by a large centrifugal barrier, which prevents the appearance of halo characteristics. As pointed out in Refs. \cite{BM2,Hamamoto2005}, open-shell nuclei with neutrons in the nearly degenerate $1f_{7/2}$ and $2p_{3/2}$ shells can induce 
a large quadrupole deformation in the ground state due to a  strong quadrupole-quadrupole interaction among nucleons in these two orbits. This may also serve as a driving mechanism for the island of inversion, where many-particle many-hole configurations are dominant in the ground state. 
The one-particle motion in a deformed potential, such as the Nilsson diagram provides key insights into understanding the microscopic structure of such nuclei. 

The halo structure induces a large concentration of
electric dipole ($E1$) strength in the low-excitation energy
region, which is referred to as soft dipole excitation \cite{NAKAMURA1997,Nakamura1999,Nakamura2006}. Recently, the
Coulomb breakup cross sections for $^{31}$Ne were measured
by Nakamura et al. \cite{Nakamura2009,Nakamura2014}, indicating a soft dipole excitation in the $^{31}$Ne nucleus. Notice that a naive spherical shell
model leads to the $1f_{7/2}$ configuration for the valence
neutron of $^{31}$Ne. In order to generate the halo structure
within the mean-field framework, the valence neutron therefore needs to move in a deformed mean-field potential, where the $s$ or $p$ wave component of a weakly-bound single-particle wave function makes a dominant contribution \cite{MISU1997,Hamamoto2012}. 

Theoretically, deformed halo nuclei have been studied within the Nilsson model  \cite{Hamamoto2005,Hamamoto2012,Hamamoto2010}, the particle-rotor model \cite{Urata2011,Urata2012,Urata2013}, the shell model \cite{Otsuka93,Kuo97}, the cluster model \cite{DESCOUVEMONT1999}, 
 the deformed relativistic Hartree-Bogoliubov theory in continuum (DRHBc) \cite{Zhou10,li12,pei131,pei132,SUN18,Nakada18,zhang19,SUN20,SUN21,SUN20212072,peisy21,ZhangKY23,ZHANG2023138112,plbzhang2024}, and anti-symmetrized molecular dynamics  (AMD) model \cite{Takatsu2023}. In this paper, we adopt a deformed Woods-Saxon model to study the intrinsic deformation effects on the dipole response of these nuclei. For largely deformed nuclei, the adiabatic approach, such as the Nilsson model and the deformed Woods-Saxon model, is adequate to examine the general features of deformation in the excitation spectrum and the transition rate. Additionally, we can easily implement the halo nature of loosely bound nucleons in the low-energy response by constraining the empirical separation energy in the deformed wave function.  

The main aim of this paper is to study how deformed halo affects the soft dipole excitation near the neutron threshold. Specifically, we will investigate the configuration dependence and the separation energy dependence of dipole response using the deformed Woods-Saxon potential. As examples, we select typical deformed halo nuclei with mass greater than $A=20$, $^{31}$Ne and $^{37}$Mg.
This paper is organized as follows: Section \ref{TF} is devoted to the basic formalism.
The results are presented in Section \ref{RD}. The summary and future perspectives are given in Section \ref{SUM}.
\section{Theoretical framework}\label{TF}

We begin with our formulation of halo wave function for a spherical shape. The neutron bound-state wave function is obtained  as  an eigenfunction of the Woods-Saxon potential with an energy eigenvalue $\varepsilon<0$
\begin{equation} \label{bound}
\left|\Phi^{(b)}: \ell j m\right\rangle=\frac{1}{r} R_{n \ell j}^{(b)}(r)\left[Y_{\ell} \otimes \chi_{1 / 2}\right]_{j m},
\end{equation} 
where the radial wave function $R_{n \ell j}^{(b)}(r)$ is a solution of Schr\"odinger equation in the  coordinate space, and  
 the normalization 
  is given by
\begin{equation}
\int_0^{\infty} d r\left|R_{n \ell j}^{(b)}(r)\right|^2  =1 .
\end{equation}
The single-neutron wave function in the continuum can be expressed in a plane wave approximation (PWA) for a  real energy variable $\varepsilon >0$ as 
\begin{equation}  \label{cwf}
\begin{aligned}
\left|\Phi_{\varepsilon}^{(c)}: \ell j m\right\rangle= & \frac{1}{r} R_{\ell j}^{(c)}(\varepsilon, r)\left[Y_{\ell} \otimes \chi_{1 / 2}\right]_{j m} \\
= & \sqrt{\frac{2 \mu}{\hbar^2 \pi k}}\left[\cos \left(\delta_{\ell j}\right) k j_{\ell}(k r)-\sin \left(\delta_{\ell j}\right) k n_{\ell}(k r)\right] \\
& \times\left[Y_{\ell} \otimes \chi_{1 / 2}\right]_{j m},
\end{aligned}
\end{equation}
where $j_{\ell}(k r)$ and $n_{\ell}(k r)$ are the spherical Bessel functions of the first kind and the second  kind, respectively, and 
\begin{equation}
k^2=\frac{2 \mu \varepsilon}{\hbar^2}, 
\end{equation}
with the reduced mass $\mu=(A-1)M/A$ for the system with the mass $A$. The phase shift is denoted as $\delta_{lj}$.
The normalization of the radial wave function is expressed as
\begin{equation}
\int_0^{\infty} d r R_{\ell j}^{(c)*}(\varepsilon, r) R_{\ell j}^{(c)}\left(\varepsilon^{\prime}, r\right)=\delta\left(\varepsilon-\varepsilon^{\prime}\right) .
\end{equation}


We consider the Hamiltonian for the deformed system,  
\begin{equation}
H=-\frac{\hbar^2}{2 \mu} \nabla^2+V\left(\boldsymbol{r}\right),
\end{equation}
where $V\left(\boldsymbol{r}\right)$ is the single-particle potential for the valence neutron interacting with the deformed core, and  $\boldsymbol{r}$ is the coordinate of the valence neutron in the intrinsic frame. We use an axially symmetric deformed Woods-Saxon potential for $V$ and expand it up to the linear order of the quadrupole deformation parameter $\beta_{2}$ as
\begin{equation}
\begin{aligned}
V\left(\boldsymbol{r}\right)= & -V_{\mathrm{WS}}\left(1-F_{ls} r_0^2(\boldsymbol{l} \cdot \boldsymbol{s}) \frac{1}{r} \frac{d}{d r}\right) f(r) \\
& +V_{\mathrm{WS}} R_0 \beta_2 \frac{d f(r)}{d r} Y_{20}\left(\hat{\boldsymbol{r}}\right) \\
\equiv & V_0(r)+V_{\mathrm{def}}\left(\boldsymbol{r}\right),
\end{aligned}
\end{equation}
where $V_0$ represents  a spherical Woods-Saxon potential together with the spin-orbit ($ls$) force and $V_{\mathrm{def}}$ is the deformed part of the potential.  $f(r)$ is given by 
\begin{equation}
\begin{aligned}
f(r) & =\frac{1}{1+\exp \left(\frac{r-R}{a}\right)}.  
\end{aligned}
\end{equation}
The values of the parameters in the potential except the depth $V_{\mathrm{WS}}$ are taken from Ref. \cite{BM2}. For example, the diffuseness $a=0.67$ fm and the radius $R=r_0A^{1/3}$ with $r_0 = 1.27$ fm.
The single-particle wave function in the deformed Woods-Saxon potential is obtained as
\begin{equation}\label{Hamil}
    H|\phi_\Omega\rangle=\varepsilon_\Omega |\phi_\Omega\rangle,
\end{equation}
where $\Omega$ represents the projection  of single-particle angular-momentum along the symmetry axis, which  is a good quantum number
in the axially symmetric potential.

The electric dipole transition $I_i^\pi, K_i^\pi\rightarrow I_f^\pi, K_f^\pi$ is expressed in the laboratory frame as \cite{BM2}
\begin{equation}  \label{B-EL}
\begin{aligned}
& B\left(E 1 (\varepsilon); I_i^\pi, K_i^\pi \rightarrow I_f^\pi, K_f^\pi\right) \\
& =\left(e_{\mathrm{eff}}^n(E 1)\right)^2\left\{\left\langle I_iK_i1K_f-K_i|I_fK_f\right\rangle \right. \\
& \quad \times\left \langle K_f^\pi (\varepsilon)| r Y_{1\nu=K_f-K_i} |K_i^\pi\right \rangle \\
& \quad+(-1)^{I_i+K_i}  \left\langle I_i-K_i1K_f+K_i|I_fK_f\right\rangle   \\
& \left.\quad \times\left \langle K_f^\pi (\varepsilon) \mid r Y_{1\nu=K_i+K_f} \mid\widetilde{K_i^\pi}\right\rangle\right\}^2.
\end{aligned}
\end{equation}
For a single-nucleon halo nucleus,  
 the initial  states can be expressed by a single-particle state $|\phi_\Omega\rangle$ in Eq. \eqref{Hamil} with $K_i=\Omega$,  and  the final state is given by the continuum wave  function \eqref{cwf} assigning $I_f=j$ and $K_f=m$.   
The effective electric charge of neutrons from the center-of-mass frame is given by
\begin{equation} \label{eff}
e_{\mathrm{eff}}^n(E 1)=\frac{N_{c}Z_{v}-Z_{c}N_v}{A},
\end{equation}
where $N_c (N_v)$ and $Z_c(Z_v)$ are neutron and proton numbers of the core (valence) particles, respectively. For the dipole transition, the second term in Eq. (\ref{B-EL}) contributes only for $K_i=1/2$ state.  However, it was shown that   the contribution of the second term in Eq. (\ref{B-EL}) is very small for [330]1/2 configuration because of cancellation between major components \cite{Hamamoto2019}. Therefore it is neglected in our calculation. 
The dipole transition strength at the  excitation energy $\omega$ in the continuum is then express as
\begin{equation}\label{Bst}
\begin{aligned}
\frac{\mathrm{d} B(E1)}{\mathrm{d} \omega}& =\int \mathrm{d} \varepsilon \delta\left(\omega-\left(\varepsilon+S\right)\right) 
\\
&\times B\left(E 1 (\varepsilon); I_i^\pi, K_i^\pi \rightarrow I_f^\pi, K_f^\pi\right),
\end{aligned}
\end{equation}
where $S>0$ is the separation energy of the last occupied neutron in the deformed Wood-Saxon potential and $\varepsilon$ is the energy of continuum state in Eq. \eqref{cwf}. 
The radial wave function in the initial wave function is expanded using spherical radial wave functions
\begin{equation}
R_{lj}(\varepsilon, r)= \sum_n c_{nlj} R_{nlj}(r).
\end{equation} 
In the present calculations, the spherical single-particle energy was truncated at 10 MeV, the orbital angular momentum was truncated at $l=10$.  In Ref. \cite{Hamamoto2010}, only the halo components, $s$ and $p$ waves, of the deformed wave functions are considered in the dipole matrix elements.  However, in the present study, all multipoles with  $l\leq 4$ are taken into account in the dipole response calculations.

We should notice that, in the following discussion, we will use the asymptotic quantum-numbers $[N n_z \Lambda]\Omega$ to denote the initial state. In the Nilsson orbit, $N$, $n_z$, $\Lambda$ and $\Omega$ represent the total oscillator quantum number, the oscillator quantum number in the $z$-direction, the orbital angular momentum projected on the symmetry axis, and the projection of total angular momentum on the symmetry axis, respectively. For example, $\left \langle K_f^\pi | r Y_{1\nu=K_f-K_i} |K_i^\pi\right \rangle$ matrix element can be expressed as $\left \langle K_f^\pi | r Y_{1\nu=K_f-K_i} |[N_i n_{zi} \Lambda_i]\Omega_i=K_i^\pi\right \rangle$.

\section{Results and discussions}\label{RD}
\subsection{Electric dipole response in $^{31}$Ne}
The measured low-excitation energies of the first 2$^+$ states of
both $^{30}$Ne \cite{Yanagisawa2003} and $^{32}$Ne 
 \cite{Door2009} are consistent with the picture that
these Ne isotopes lie inside the
island of inversion. Moreover, the large Coulomb breakup
cross section reported in Refs. \cite{Nakamura2009,Nakamura2014}, which clearly indicates the halo nature of the ground state of $^{31}$Ne, suggests a substantial contribution from the $p$ component of the 21st
neutron in the deformed mean field.  
It is also noted that the measured spin and magnetic moment of the ground state of $^{33}$Mg, which has the same neutron number $N = 21$ as $^{31}$Ne, are reported to be  consistent with the interpretation of $I^{\pi}= \frac{3}{2}^-$ in Ref. \cite{Yordanov2007}. 

Low-lying states of odd-$A$ medium-heavy deformed nuclei
are well approximated by one (quasi)particle
moving in the deformed potential produced by the even-even
core \cite{BM2}. The single-particle picture works better in deformed nuclei than in spherical nuclei, because the major part of the long-range residual interaction  can
be included in the deformed mean field, and consequently in the deformed wave function.  The halo nature will be implemented in 
these wave functions by solving the Schr\"odinger equation in the coordinate space.
The harmonic oscillator wave functions are often used
in the standard shell model calculations,  but their applicability to the description of halo nuclei is questionable  \cite{zhou03}.  

\begin{figure}
 \centering
 \includegraphics[width=8cm]{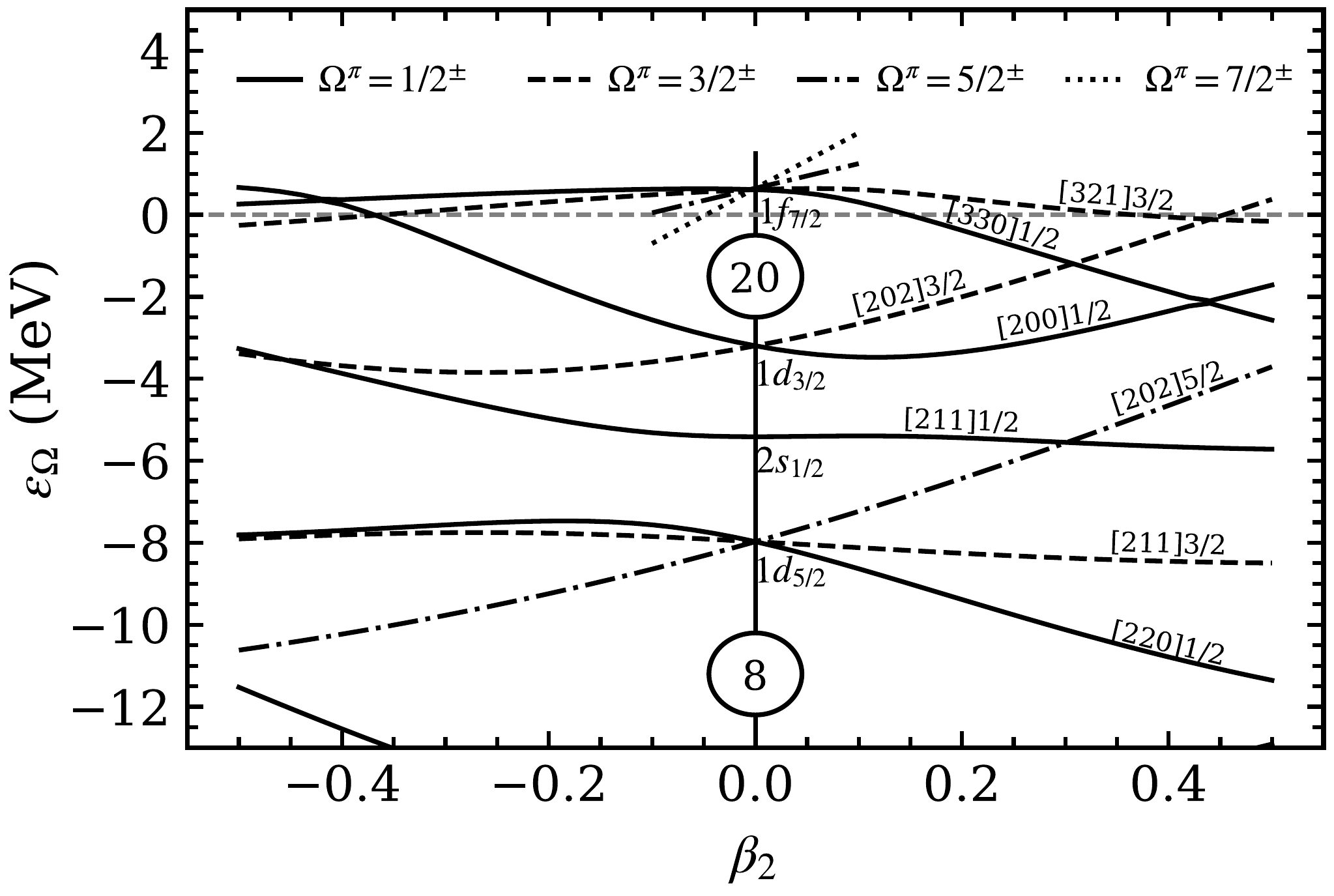}
 \caption{Single-particle levels for neutrons in deformed Woods-Saxon potentials as a function of the quadrupole deformation parameter $\beta_{2}$. The potential
depth $V_{\mathrm{WS}}$ is adjusted so that 
the binding energy of the 21st neutron of the prolate deformed nucleus $^{31}$Ne is 150 keV.
The asymptotic quantum-numbers $[Nn_z\Lambda]\Omega$ are denoted for the single particle levels. 
}   \label{fig1}
\end{figure}  

Figure \ref{fig1} shows the Nilsson diagram calculatd by using deformed Woods-Saxon potential, whose  
parameters are adjusted for the separation energy of $S_n\sim150$ keV of  ($^{30}\text{Ne}+n$)
system.
The one neutron outside the $N=20$ core occupies the negative parity $[330]1/2$ orbit in the small deformation region $0.15<\beta_2<0.3$, while it occupies $[321]3/2$ orbit at very large deformation $\beta_2>0.4$. In the medium deformation range, around $0.3<\beta_2<0.4$, the last neutron occupies a positive parity $[202]3/2$ orbit. As shown in Tables \ref{tab1}, \ref{tab2} and \ref{tab3}, $p_{3/2}$, $f_{5/2}$ and $f_{7/2}$ are included in the calculation of the $[321]3/2$,  while the $p_{1/2}$, $p_{3/2}$, $f_{5/2}$ and $f_{7/2}$ are included in the calculation of the$[330]1/2$. In the calculation of the $[202]3/2$, the $d_{3/2}$ and $d_{5/2}$ orbits are included. It should be noted that the negative parity orbits in Tables  \ref{tab1} and \ref{tab2} have significant contributions from the halo $p$-components in the deformed wave function, while the positive parity states in Table \ref{tab3}  have  no halo component.  The probability of $p_{3/2}$ component in [321]3/2 configuration is  32.7\% which is almost identical  to the value of 32\% obtained by the Coulomb breakup cross section experiment in Ref. \cite{Nakamura2014}.

\begin{table}[ht!]  
\centering
\caption{The decomposition of the matrix elements $\langle K_f^\pi=\frac{3}{2}^+|rY_{10}|[321]\rangle$, $\langle K_f^\pi=\frac{1}{2}^+|rY_{1-1}|[321]3/2\rangle$ and $\langle K_f^\pi=\frac{5}{2}^+|rY_{11}|[321]3/2\rangle$ in $^{31}$Ne by contributions of  spherical wave functions at the peak energy of the response $\varepsilon_{\rm peak}$. The unit is fm. }\label{tab1}
\begin{tabular}{c>{\centering\arraybackslash} p{0.7cm}>{\centering\arraybackslash} p{1.27cm}>{\centering\arraybackslash} p{1.27cm}>{\centering\arraybackslash} p{1.27cm}c}
\hline
\hline
  Initial state &  & \multicolumn{3}{c}{ ($[321]3/2$, $I_i^\pi=\frac{3}{2}^-$,  $K_i^\pi=\frac{3}{2}^-$) }  \\
\multicolumn{2}{c}{Spherical configuration}  & $p_{3/2}$ & $f_{5/2}$& $f_{7/2}$\\
Probability  & & 0.327&$0.022$&$0.650$\\\hline
\multirow{5}{*}{\makecell{$I_f^\pi=\frac{1}{2}^+,\frac{3}{2}^+,\frac{5}{2}^+$\\ \\$K_f^\pi=\frac{1}{2}^+$ \\ \\$\varepsilon_{\rm peak}=0.07$ MeV} } &	$s_{1/2}$	&$	-2.160 	$&$	      $&$		$\\	
&	$d_{3/2}$	&$	-0.070 	$&$	-0.002 	$&$		$\\
	&	$d_{5/2}$	&$	~~0.085 	$&$	-0.001 	$&$	-0.040 	$\\
	&	$g_{7/2}$	&$		$&$	~~0.000 	$&$	~~0.000 	$\\
	                            &	$g_{9/2}$	&$		$&$		$&$	~~0.001 	$\\\hline
\multirow{4}{*}{\makecell{$I_f^\pi=\frac{3}{2}^+,\frac{5}{2}^+$\\ \\$K_f^\pi=\frac{3}{2}^+$\\ \\$\varepsilon_{\rm peak}=0.57$ MeV} }	&	$d_{3/2}$	&$	-0.339 	$&$	-0.050 	$&$		$\\
	&	$d_{5/2}$	&$	~~0.678 	$&$	~~0.011 	$&$	~~0.200 	$\\
	&	$g_{7/2}$	&$		$&$	~~0.000 	$&$	-0.006 	$\\
	&	$g_{9/2}$	&$		$&$		$&$	~~0.059 	$\\
\hline									
\multirow{3}{*}{\makecell{$I_f^\pi=\frac{5}{2}^+, K_f^\pi=\frac{5}{2}^+$\\ \\$\varepsilon_{\rm peak}=0.50$ MeV} }	&	$d_{5/2}$	&$	~~1.087 	$&$	-0.010 	$&$	-0.061 	$\\
	&	$g_{7/2}$	&$		$&$	~~0.004 	$&$	~~0.008 	$\\
	&	$g_{9/2}$	&$		$&$		$&$	~~0.055 	$\\
\hline
\hline
\end{tabular}
\end{table}

\begin{table}[ht!]  
\centering
\caption{The decomposition of 
the matrix elements $\langle K_f^\pi=\frac{1}{2}^+|rY_{10}|[330]1/2\rangle$ and $\langle K_f^\pi=\frac{3}{2}^+|rY_{11}|[330]1/2\rangle$ in $^{31}$Ne by contributions of  spherical wave functions at the peak energy of the response $\varepsilon_{\rm peak}$. The unit is fm. } \label{tab2}
\begin{tabular}{cccccc}
\hline
\hline
 Initial state & & \multicolumn{4}{c}{ ($[330]1/2$, $I_i^\pi=\frac{3}{2}^-$, $K_i^\pi=\frac{1}{2}^-$)}    \\
\multicolumn{2}{c}{Spherical configuration}   &$p_{1/2}$& $p_{3/2}$ & $f_{5/2}$& $f_{7/2}$\\
Probability & & $0.063$ &0.590&0.004&$0.342$\\\hline
\multirow{5}{*}{\makecell{$I_f^\pi=\frac{1}{2}^+,\frac{3}{2}^+,\frac{5}{2}^+$\\ \\$K_f^\pi=\frac{1}{2}^+$ \\ \\$\varepsilon_{\rm peak}=0.09$ MeV} }	&	$s_{1/2}$	&\multicolumn{1}{r}{$	-0.640 	$}&\multicolumn{1}{r}{$	-2.650 	$}&\multicolumn{1}{r}{$		$}&\multicolumn{1}{r}{$		$}\\
	&	$d_{3/2}$	&\multicolumn{1}{r}{$	0.136 	$}&\multicolumn{1}{r}{$	0.057 	$}&\multicolumn{1}{r}{$	-0.001 	$}&\multicolumn{1}{r}{$		$}\\
	&	$d_{5/2}$	&\multicolumn{1}{r}{$		$}&\multicolumn{1}{r}{$	-0.421 	$}&\multicolumn{1}{r}{$	0.000 	$}&\multicolumn{1}{r}{$	-0.035 	$}\\
	&	$g_{7/2}$	&\multicolumn{1}{r}{$		$}&\multicolumn{1}{r}{$		$}&\multicolumn{1}{r}{$	-0.001 	$}&\multicolumn{1}{r}{$	0.000 	$}\\
	&	$g_{9/2}$	&\multicolumn{1}{r}{$		$}&\multicolumn{1}{r}{$		$}&\multicolumn{1}{r}{$		$}&\multicolumn{1}{r}{$	-0.001 	$}\\\hline
\multirow{4}{*}{\makecell{$I_f^\pi=\frac{3}{2}^+,\frac{5}{2}^+$ \\ \\$K_f^\pi=\frac{3}{2}^+$\\ \\$\varepsilon_{\rm peak}=0.49$ MeV} }	&	$d_{3/2}$	&\multicolumn{1}{r}{$	0.514 	$}&\multicolumn{1}{r}{$	-0.423 	$}&\multicolumn{1}{r}{$	-0.010 	$}&\multicolumn{1}{r}{$		$}\\
	&	$d_{5/2}$	&\multicolumn{1}{r}{$		$}&\multicolumn{1}{r}{$	-1.270 	$}&\multicolumn{1}{r}{$	0.006 	$}&\multicolumn{1}{r}{$	0.075 	$}\\
	&	$g_{7/2}$	&\multicolumn{1}{r}{$		$}&\multicolumn{1}{r}{$		$}&\multicolumn{1}{r}{$	-0.002 	$}&\multicolumn{1}{r}{$	-0.005 	$}\\
	&	$g_{9/2}$	&\multicolumn{1}{r}{$		$}&\multicolumn{1}{r}{$		$}&\multicolumn{1}{r}{$		$}&\multicolumn{1}{r}{$	-0.026 	$}\\
\hline
\hline
\end{tabular}
\end{table}


\begin{figure}
 \centering
 \includegraphics[width=8cm]{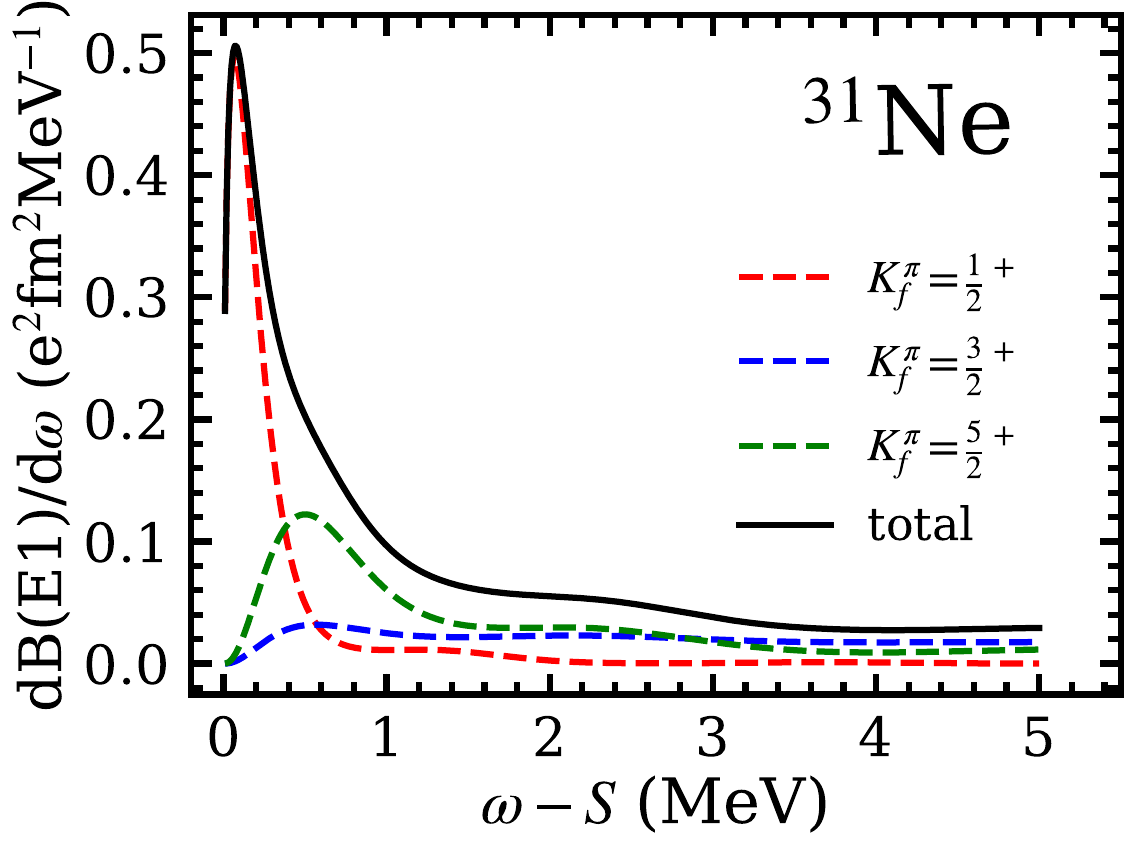}
 \caption{ (Color online) Dipole strength distribution calculated in PWA as function of the excitation energy (referred to the particle threshold). The wave function of the ground state takes the $[321]3/2, I^\pi_i=\frac{3}{2}^-$ configuration at $\beta_{2}=0.5$, which is calculated using the deformed Woods-Saxon potential, corresponding the single-neutron separation energy of $0.15$ MeV.  The $p_{3/2}$, $f_{5/2}$ and $f_{7/2}$ orbitals are included in the calculation of the $[321]3/2$ orbit.  All final angular momenta $I^\pi_f$ are summed up for each $K^\pi_f$ in the figure. 
}   \label{fig2}
\end{figure}  

Due to the selection rules of the transition operator for calculating the $E1$ response, the final states  can be $s$-wave, $d$-wave and $g$-wave. The configuration $[321]3/2$ is taken as the initial state with $I_i^\pi=K_i^\pi=\frac{3}{2}^-$, and the corresponding $I_f^\pi$ and $K_f^\pi$ are shown in Table \ref{tab1}.  
Using  Eqs. (\ref{B-EL})-(\ref{Bst}), we calculate the dipole strength distribution from the $[321]3/2$ configuration to six different final states, as well as the total strength distribution. The results are shown in Fig. \ref{fig2}, where the final angular momenta $I^\pi_f$ are summed up for each $K^\pi_f$.
It can be seen that the $E1$ response to the $K^\pi_f=\frac{1}{2}^+$ state shows a very sharp peak just above the threshold, while the $K^\pi_f=\frac{3}{2}^+$ and $\frac{5}{2}^+$ states show relatively broad peaks at location slightly higher than the threshold. The contributions of these six states lead to an enhanced single-peak structure in the $E1$ strength distribution near the threshold. 

\begin{figure} 
\centering
\includegraphics[width=8cm]{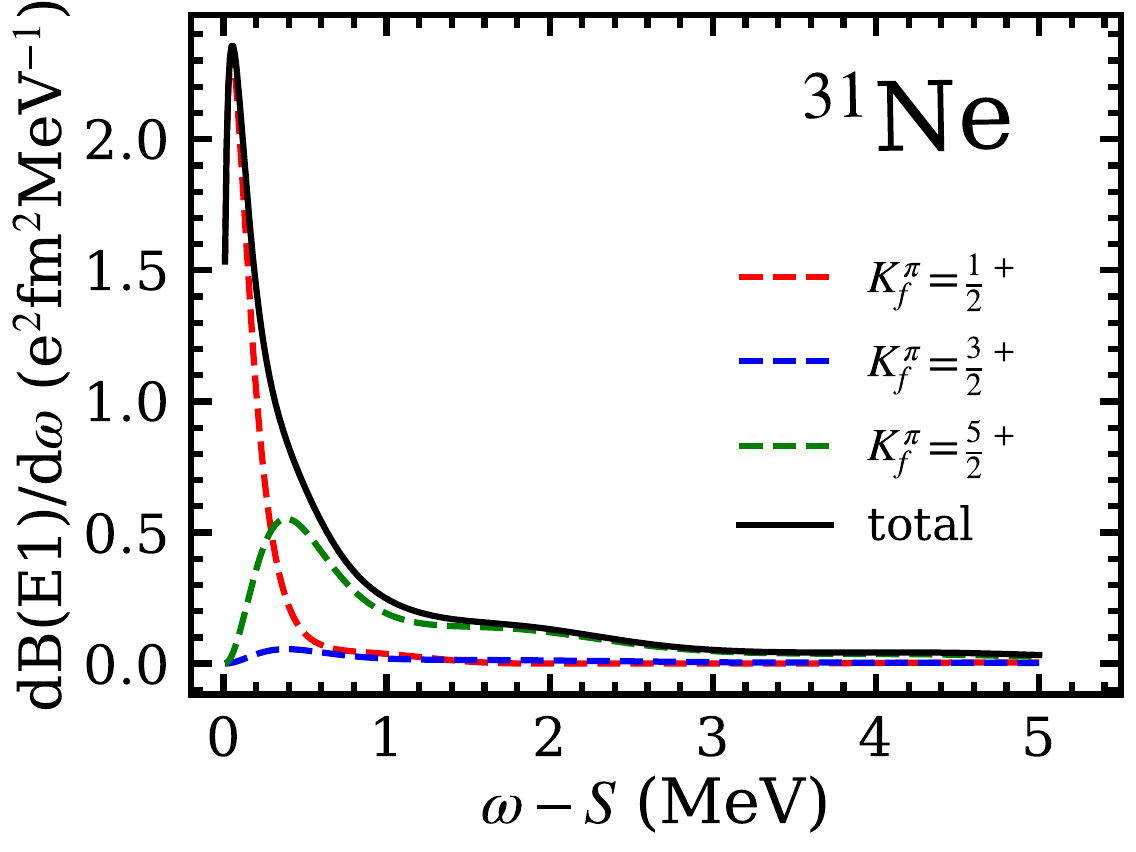}
\caption{  (Color online) Same as Fig. \ref{fig2}, but for 
a pure single-particle halo $2p_{3/2}$ state  taken from $[321]3/2$ configuration with the separation energy $S=0.15$ MeV.
}  \label{fig3}
\end{figure}

The integrated strengths up to $\omega -S=5$ MeV are listed in Table \ref{tab2}. The total transition strength is much larger than the low-energy $E1$ transition in non-halo nuclei, which range from $10^{-3}\sim 10^{-4}$ $e^2$fm$^2$, but is smaller than the pure single-particle transition from a halo $p$-wave state having  the same  separation energy $S_n=0.15$ MeV as  the [321]3/2 configuration, as shown in Fig. \ref{fig3}. 
Since the responses to the configurations of   $K_f^{\pi}=\frac{3}{2}^+$ and $K_f^{\pi}=\frac{5}{2}^+$ have no $s$-wave component of wave functions in the final states, they contribute to the dipole response with  slightly higher energy and larger width than the sharp peak of 
$K_f^{\pi}=\frac{1}{2}^+$ configuration  at $\varepsilon\sim0.07$ MeV above the threshold. 
From Table \ref{tab1}, it can be seen that this is due to 32.7\% contribution from the $p_{3/2}$ component, along with a portion of the matrix elements from $f$-wave scattering to $d$- and $g$-waves that partially cancel the contribution to $E1$ response.

\begin{figure}
 \centering
 \includegraphics[width=8cm]{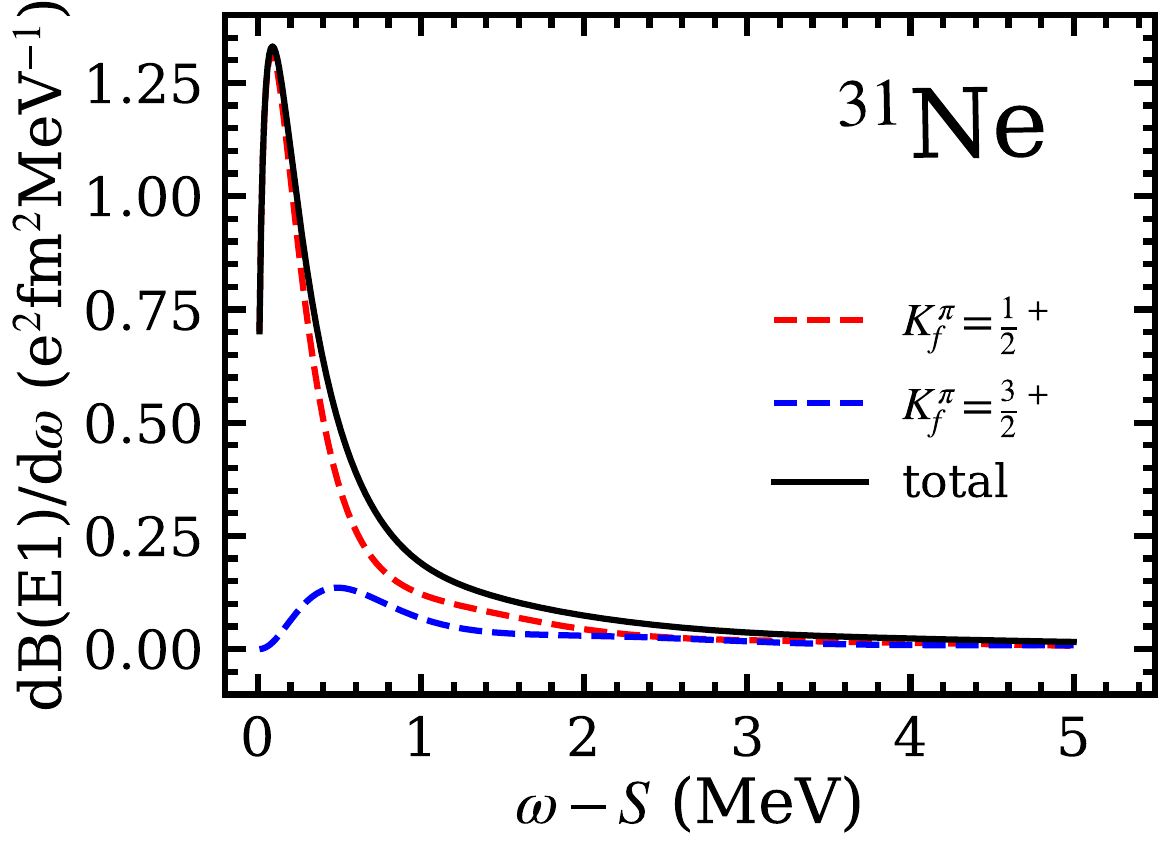}
 \caption{ (Color online) Same as Fig. \ref{fig2}, but for the wave function of the ground state $[330]1/2, I_i^\pi=\frac{3}{2}^-$ at $\beta_{2}=0.24$. The $p_{1/2}$, $p_{3/2}$, $f_{5/2}$ and $f_{7/2}$ orbitals are included in the calculation of the $[330]1/2$ orbit.
}   \label{fig4}
\end{figure}

The $E1$ response from the initial state $[330]1/2$ at $\beta_{2}=0.24$ is shown in Fig. \ref{fig4}. For this configuration, the final states are limited to the states with $K_f^\pi=\frac{1}{2}^+$ and $ \frac{3}{2}^+$. Again, the transitions to $K^{\pi}_f=\frac{1}{2}^+$ exhibit a sharp peak just above the threshold, while the $E1$ response to $K^\pi_f=\frac{3}{2}^+$ states shows a relatively broad peak at a location slightly higher than the threshold. We should note that the total $E1$ strength of the $[330]1/2$ configuration is nearly 2 times that  of the $[321]3/2$ state, as shown in Table \ref{tab4}. This is because the halo $p$-wave component in the $[330]1/2$ state in Table \ref{tab2} is almost twice as large as that in the $[321]3/2$ state.  
 
\begin{figure}
 \centering
 \includegraphics[width=8cm]{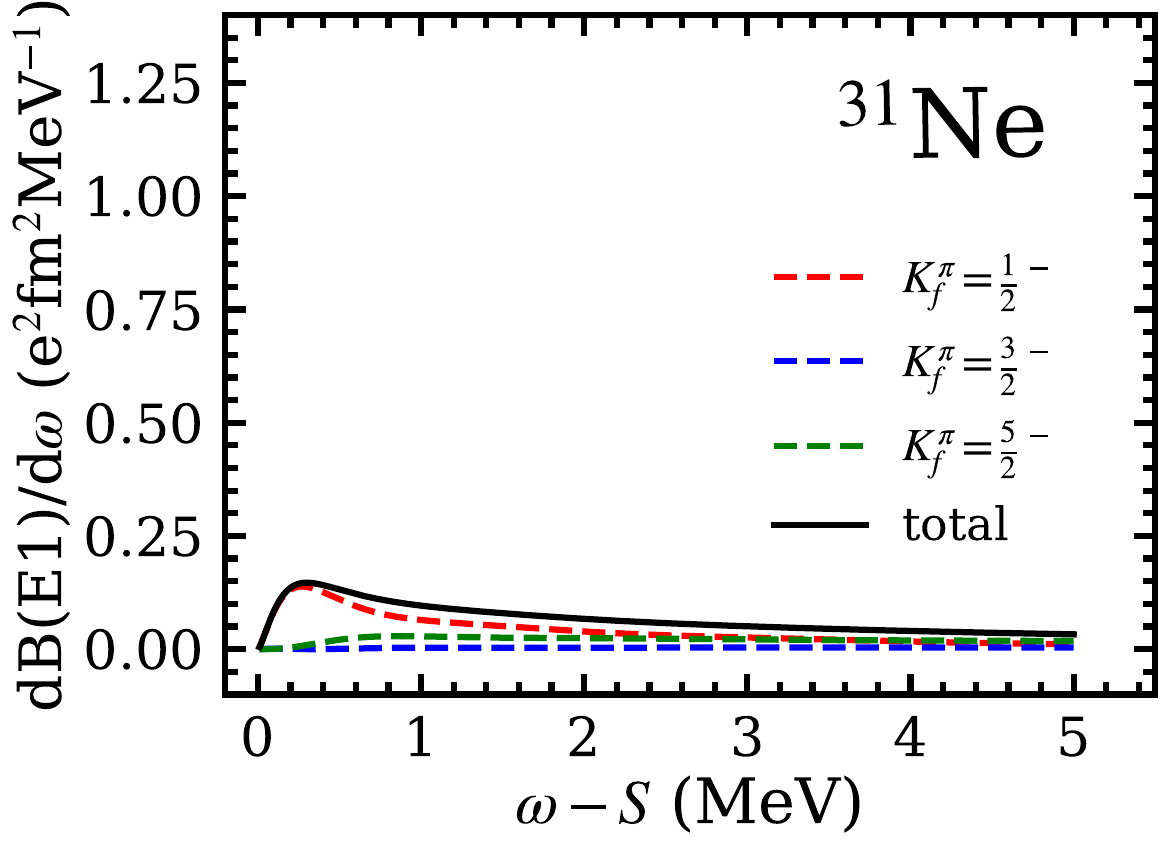}
 \caption{ (Color online) Same as Fig. \ref{fig2}, but for the  different wave function of the ground state $[202]3/2, I_i^\pi=\frac{3}{2}^+$ at $\beta_{20}=0.32$. The $d_{3/2}$ and $d_{5/2}$ orbitals are included in the calculation of the $[202]3/2$ orbit.
}   \label{fig5}
\end{figure}  

The $E1$ response from $[202]3/2$ state is shown in Fig. \ref{fig5}.
This positive parity configuration appears to be a possible $N=21$ configuration in the medium deformation range at $\beta_2\sim0.3$. The dipole response shows a broad peak above the threshold, as there is no component of the halo $s$-wave in this Nilsson orbit, as tabulated in Table \ref{tab3}.  The integrated $E1$ strength, tabulated in Table \ref{tab4}, is also relatively smaller than those for the configurations $[330]1/2$ and $[321]3/2$. From the dipole response discussed here,   it could be  possible to  distinguish the configuration of the last neutron in $^{31}$Ne by looking at  the enhancement of peak height, the  width  of peak and the integrated  dipole strength distribution as shown in Table \ref{tab4}. 

\begin{table}[ht!]  
\centering
\caption{The decomposition of 
the matrix elements 
$\langle K_f^\pi=\frac{3}{2}^- |rY_{10}|[202]3/2\rangle$, $\langle K_f^\pi=\frac{1}{2}^-|rY_{1-1}|[202]3/2\rangle$ and $\langle K_f^\pi=\frac{5}{2}^-|rY_{11}|[202]3/2\rangle$
in $^{31}$Ne by contributions of  spherical wave functions at the peak energy of the response $\varepsilon_{\rm peak}$. The unit is fm. } \label{tab3}
\begin{tabular}{c>{\centering\arraybackslash} p{1cm}>{\centering\arraybackslash}p{1.5cm}>{\centering\arraybackslash}p{1.5cm}cccc}
\hline
\hline
Initial state &   \multicolumn{3}{c}{($[202]3/2$, 
 $I_i^\pi=\frac{3}{2}^+, K_i^\pi=\frac{3}{2}^+$)}  &    \\
Spherical configuration  & &  $d_{3/2}$ & $d_{5/2}$ \\
Probability & & $0.965$ & $0.034$\\\hline 
\multirow{4}{*}{\makecell{$I_f^\pi=\frac{1}{2}^-,\frac{3}{2}^-,\frac{5}{2}^-$ \\ \\$K_f^\pi=\frac{1}{2}^-$ \\ \\$\varepsilon_{\rm peak}=0.27$ MeV} }&	$p_{1/2}$	&$	-0.865 	$&$		$\\
&	$p_{3/2}$	&$	-0.245 	$&$	-0.124 	$\\
&	$f_{5/2}$	&$	~~0.071 	$&$	-0.007 	$\\
&	$f_{7/2}$	&$		$&$	~~0.015 	$\\\hline
\multirow{3}{*}{\makecell{$I_f^\pi=\frac{3}{2}^-,\frac{5}{2}^-$\\$K_f^\pi=\frac{3}{2}^-$\\ $\varepsilon_{\rm peak}=4.6$ MeV} }&	$p_{3/2}$	&$	-0.115 	$&$	~~0.041 	$\\
&	$f_{5/2}$	&$	~~0.230 	$&$	-0.008 	$\\
&	$f_{7/2}$	&$		$&$	~~0.044 	$\\\hline
$I_f^\pi=\frac{5}{2}^-$, $K_f^\pi=\frac{5}{2}^-$  &	$f_{5/2}$	&$	~~0.450 	$&$	~~0.011 	$\\
$\varepsilon_{\rm peak}=0.87$ MeV&	$f_{7/2}$	&$		$&$	~~0.067 	$\\
\hline
\hline
\end{tabular}
\end{table}


\begin{table*}[ht!]
\centering
\caption{ Dipole transition probabilities $B(E1)$ ($e^2$fm$^2$) for $^{31}$Ne with different initial wave functions.}  \label{tab4}
\begin{tabular}{ccccccccccc}
\hline
\hline
\multirow{2}{*}{Initial state} & $[321]3/2$ &  $[330]1/2$ & $[202]3/2$  & pure $2p_{3/2}$ \\
 &($I_i^\pi=K_i^\pi=\frac{3}{2}^-$) &  ($I_i^\pi=\frac{3}{2}^-$, $K_i^\pi=\frac{1}{2}^-$) & ($I_i^\pi=K_i^\pi=\frac{3}{2}^+$)&($I_i^\pi=K_i^\pi=\frac{3}{2}^-$) \\
\hline
$K_f^\pi=\frac{1}{2}^+$&	0.147	& 0.655 & 0.212 & 0.519	\\ 
$K_f^\pi=\frac{3}{2}^+$&	0.103	& 0.176 & 0.016	& 0.064\\ 
$K_f^\pi=\frac{5}{2}^+$&	0.164	& --- & 0.106	& 0.638\\ \hline
Total&	0.414 	& 	0.831  &	0.334 	& 1.221 \\\hline
\hline
\end{tabular}
\end{table*}
In Ref. \cite{Hamamoto2010}, the large Coulomb dissociation cross sections of $^{31}$Ne was discussed by using a deformed Woods-Saxon potential with three possible configurations of $[Nn_z\Lambda]\Omega=[330]1/2$, $[321]3/2$, or $[200] 1/2$.
We note that the ground state with $I^\pi= \frac{3}{2}^-$ is predicted in the shell model calculations \cite{POVES1994}, as well as in the microscopic cluster model calculations \cite{DESCOUVEMONT1999} and in AMD calculations \cite{Takatsu2023}.  A decisive experimental information could be obtained from the soft dipole excitation near the neutron threshold. 

 \begin{table}[ht!]  
\centering
\caption{The decomposition of 
the matrix elements $\langle K_f^\pi=\frac{1}{2}^+|rY_{10}|[321]1/2\rangle$ and $\langle K_f^\pi=\frac{3}{2}^+|rY_{11}|[321]1/2\rangle$ in $^{37}$Mg by contributions of spherical wave functions at the peak energy of the response $\varepsilon_{\text{peak}}$. The unit is fm.} \label{tab5}
\begin{tabular}{cccccc}
\hline
\hline
Initial configuration & &   \multicolumn{4}{c}{ $[321]1/2$~ ($I_i^\pi=\frac{3}{2}^-$, $K_i^\pi=\frac{1}{2}^-$)}    \\
\multicolumn{2}{c}{Spherical configuration} &  $p_{1/2}$&$p_{3/2}$&$f_{5/2}$ & $f_{7/2}$ \\
Probability & & 0.428&$0.149$& $0.077$ & $0.345$\\\hline  
\multirow{5}{*}{\makecell{$I_f^\pi=\frac{1}{2}^+,\frac{3}{2}^+,\frac{5}{2}^+$\\ \\$K_f^\pi=\frac{1}{2}^+$\\ \\$\varepsilon_{\rm peak}=0.06$ MeV} }	&	$s_{1/2}$	&\multicolumn{1}{r}{$	1.438 	$}&\multicolumn{1}{r}{$	1.406 	$}&\multicolumn{1}{r}{$		$}&\multicolumn{1}{r}{$		$}\\
	&	$d_{3/2}$	&\multicolumn{1}{r}{$	-0.181 	$}&\multicolumn{1}{r}{$	-0.019 	$}&\multicolumn{1}{r}{$	-0.010 	$}&\multicolumn{1}{r}{$		$}\\
	&	$d_{5/2}$	&\multicolumn{1}{r}{$		$}&\multicolumn{1}{r}{$	0.143 	$}&\multicolumn{1}{r}{$	0.001 	$}&\multicolumn{1}{r}{$	-0.020 	$}\\
	&	$g_{7/2}$	&\multicolumn{1}{r}{$		$}&\multicolumn{1}{r}{$		$}&\multicolumn{1}{r}{$	0.000 	$}&\multicolumn{1}{r}{$	0.000 	$}\\
	&	$g_{9/2}$	&\multicolumn{1}{r}{$		$}&\multicolumn{1}{r}{$		$}&\multicolumn{1}{r}{$		$}&\multicolumn{1}{r}{$	-0.001 	$}\\\hline
\multirow{4}{*}{\makecell{$I_f^\pi=\frac{3}{2}^+,\frac{5}{2}^+$\\ \\$K_f^\pi=\frac{3}{2}^+$\\ \\$\varepsilon_{\rm peak}=0.67$ MeV} }	 &	$d_{3/2}$	&\multicolumn{1}{r}{$	-1.074 	$}&\multicolumn{1}{r}{$	0.199 	$}&\multicolumn{1}{r}{$	0.046 	$}&\multicolumn{1}{r}{$		$}\\
	&	$d_{5/2}$	&\multicolumn{1}{r}{$		$}&\multicolumn{1}{r}{$	0.598 	$}&\multicolumn{1}{r}{$	-0.026 	$}&\multicolumn{1}{r}{$	0.070 	$}\\
	&	$g_{7/2}$	&\multicolumn{1}{r}{$		$}&\multicolumn{1}{r}{$		$}&\multicolumn{1}{r}{$	-0.023 	$}&\multicolumn{1}{r}{$	-0.007 	$}\\
	&	$g_{9/2}$	&\multicolumn{1}{r}{$		$}&\multicolumn{1}{r}{$		$}&\multicolumn{1}{r}{$		$}&\multicolumn{1}{r}{$	-0.035 	$}\\
\hline
\hline
\end{tabular}
\end{table}

\subsection{Electric dipole response in $^{37}$Mg}
The observed excitation energies of neighboring nuclei $^{36}$Mg \cite{mg36-2,mg36-mg38} and $^{38}$Mg \cite{mg36-mg38} suggest that these neutron-rich magnesium isotopes lie 
within the island of inversion and indicate a strong quadrupole of deformation for $^{37}$Mg.  The parallel momentum distributions of the  $^{37}$Mg residues after  the Coulomb breakup reactions  also support the presence of strong deformation \cite{mg37}. Theoretically, strong deformation is predicted in Refs. \cite{REN1996241, Chen2005, li12, Xiong_2016, ZHANG2023138112} within the framework of relativistic and non-relativistic mean-field theories.

\begin{figure}
 \centering
 \includegraphics[width=8cm]{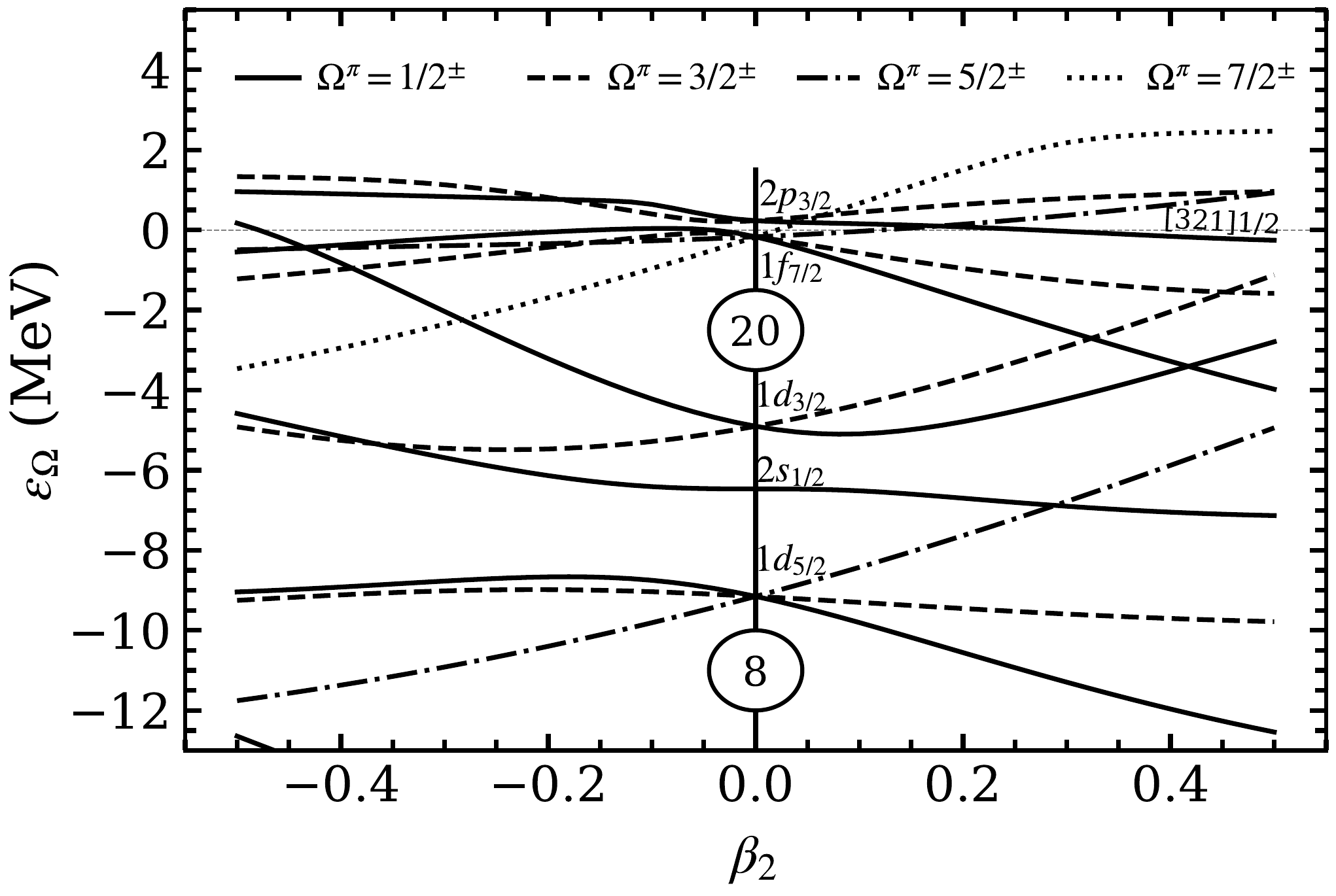}
 \caption{ Same as Fig. \ref{fig1}, but for $^{37}$Mg. The asymptotic quantum-numbers $[Nn_z\Lambda]\Omega$ represent the single-particle level, which is particularly important in the present subject.
}   \label{fig6}
\end{figure}  

The Nilsson diagram of $^{37}$Mg is shown in Fig. \ref{fig6}, where the deformed Woods-Saxon potential is adjusted for the separation energy of last neutron to be $S_n\sim0.2$ MeV. As can be seen in the figure, the last neutron, with $N=25$, may occupy the $1f_{7/2}$ orbit in the spherical limit. For the prolate deformation, the negative parity $[321]1/2$ orbit becomes the last occupied orbit for neutrons when $\beta_2>0.2$. From Table \ref{tab5}, we can see that the deformed wave function of  $[321]1/2$ has the components of $p_{1/2}$, $p_{3/2}$, $f_{5/2}$ and $f_{7/2}$ orbits, and indicates a large halo contribution of $p$-wave  with 58\% probability.

\begin{figure}
 \centering
 \includegraphics[width=8cm]{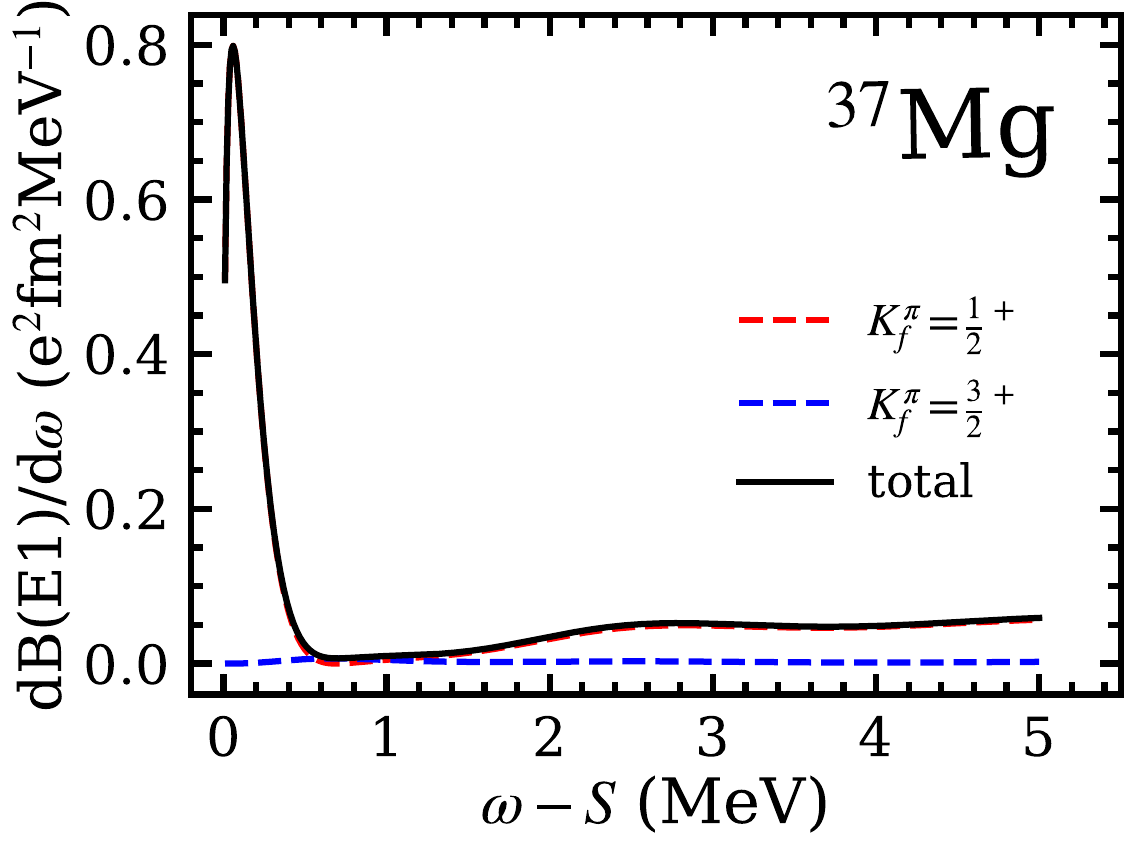}
 \caption{ (Color online) Dipole strength distribution in  $^{37}$Mg calculated as a function of the excitation energy (referred to the particle threshold). The wave function of the ground state takes the $[321]1/2, I^\pi_i=3/2^-$ configuration at $\beta_{2}=0.46$, which is calculated using the deformed Woods-Saxon potential, corresponding to the single-neutron separation energy of $0.22$ MeV.  The $p_{1/2}$, $p_{3/2}$, $f_{5/2}$ and $f_{7/2}$ orbitals are included in the calculation of the $[321]1/2$ orbit.
} \label{fig7}
\end{figure} 

Figure \ref{fig7} shows the $E1$ response from a halo configuration $[321]1/2$ with $I_i^\pi=\frac{3}{2}^-$, $K^\pi_i=\frac{1}{2}^-$ to the positive parity states with $K_f^\pi=\frac{1}{2}^+$ and $K_f^\pi=\frac{3}{2}^+$. Similar to the halo configuration $[330]1/2$ in $^{31}$Ne,
 the excitation to the $I^\pi_f=\frac{1}{2}^+$, $\frac{3}{2}^+$ and $\frac{5}{2}^+$ with $K^{\pi}_f=\frac{1}{2}^+$
 dominates the $E1$ transition strength, while there is almost no strength for the states with 
 $K_f^\pi=\frac{3}{2}^+$.  Since the transition matrix elements for $K_f^\pi=\frac{1}{2}^+$ have the same sign for the initial $p_{3/2}$ and $p_{1/2}$ configuration in Table \ref{tab5}, the dipole response becomes large for the $K_f^\pi=\frac{1}{2}^+$ state.  On the other hand, the transition matrix elements have opposite signs for $K_f^\pi=\frac{3}{2}^+$, which induce a large cancellation for the dipole response, as shown in Fig. \ref{fig7}. The dip at $\omega-S =$ 0.7 MeV is also due to the cancellation among the matrix elements of the transition to  $K_f^\pi=1/2^+$ state.  The $E1$ response increases up to 2 MeV, then decreases after 5 MeV and goes to zero eventually.

\begin{figure}
\centering
\includegraphics[width=8cm]{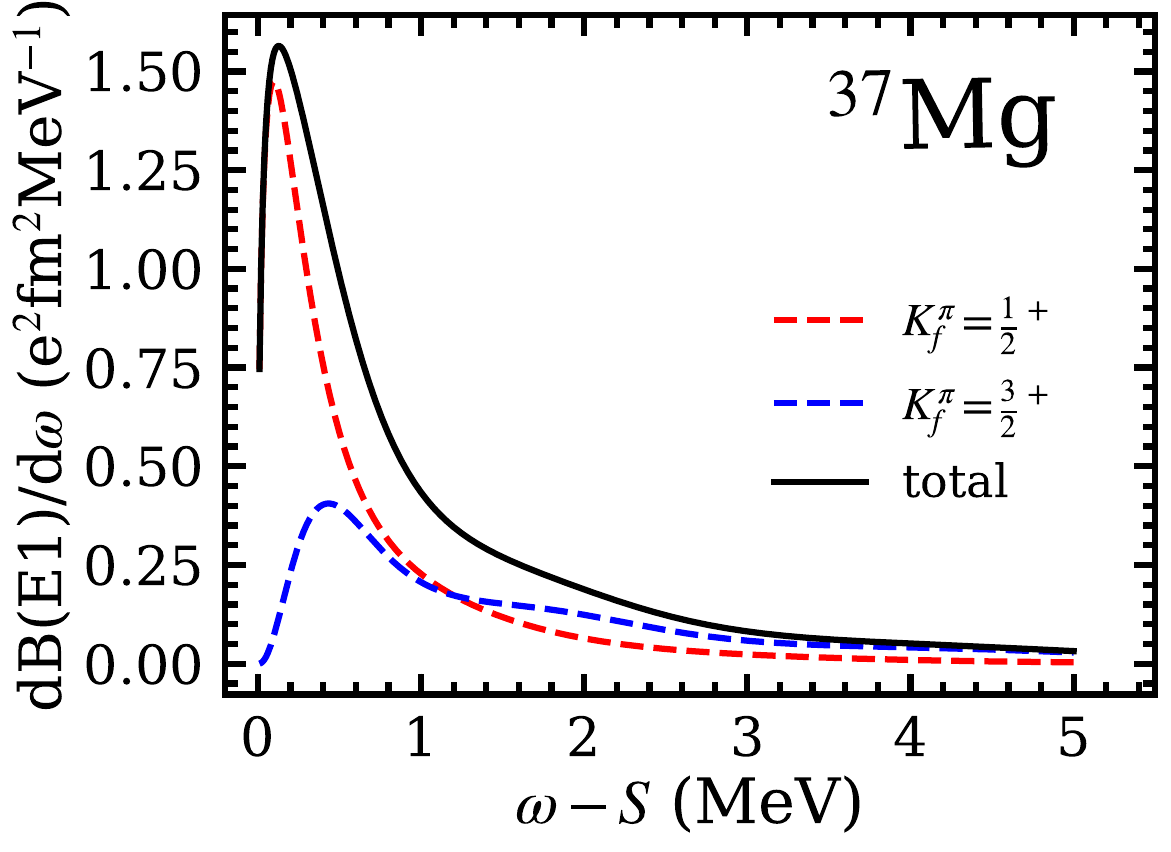}
\caption{  (Color online) Same as Fig. \ref{fig7}, but for the 
 pure $2p_{3/2}$ configuration with the separation energy $S_n=0.22$ MeV.
} \label{fig8}
\end{figure}

The dipole response with the pure halo $2p_{3/2}$ configuration with the separation energy $S_n=0.22$ MeV is shown in Fig. \ref{fig8}. The $E1$ strength in Fig. \ref{fig8} is a factor  2 larger than  the response to the deformed wave function  in Fig. \ref{fig7},  which is caused by 58\% probability of  halo $p$-wave component  of [321]1/2 orbit. The integrated $E1$ strength is listed in Table \ref{tab6}. Because of large contributions from $s$-wave and $d$-wave component of wave functions in the final states, the pure $2p_{3/2}$ configuration has almost factor 5 larger than the deformed configuration [321]1/2.

 

\begin{table}[ht!]
\centering
\caption{\label{tab6}Dipole transition probabilities $B(E1)$ ($e^2$fm$^2$) for $^{37}$Mg with different initial wave functions. 
The initial state is taken as $I^\pi_i=\frac{3}{2}^-$ and $K^\pi_i=\frac{1}{2}^-$.}  
\begin{tabular}{>{\centering\arraybackslash}p{2cm} >{\centering\arraybackslash}p{2cm}>{\centering\arraybackslash}p{2cm}}
\hline
\hline
Initial state  & $[321]1/2$ 
&  $2p_{3/2}$\\
\hline
$K_f^\pi=\frac{1}{2}^+$&	0.333	& 0.896	\\ 
$K_f^\pi=\frac{3}{2}^+$&	0.013	& 0.601  \\  \hline
Total&	0.346 	& 	1.497\\\hline
\end{tabular}
\end{table}

\section{Summary}\label{SUM}
We studied the $E1$ response from the deformed halo orbits in $^{31}$Ne and $^{37}$Mg taking into account all possible configurations  in  the Nilsson orbits, whose  separation energy is adjusted to the experimental  one. There are  three possible configurations of the ground state in $^{31}$Ne, i.e., the  Nilsson orbits, $[321]3/2$, $[330]1/2$ and $[202]3/2$  depending on the magnitude of $\beta_2$ value. We pointed out that the two halo configurations, $[321]3/2$ and $[330]1/2$, contain large amplitudes of halo $p$-wave components, which significantly enhance the threshold strength. On the other hand,   the non-halo configuration $[202]3/2$  have some enhancement of the soft dipole  strength in  the integrated $E1$ value in Table \ref{tab4}.  However, the peak does not show any sign of enhancement because of much  broader width than the halo configurations.  
 The two halo configurations, $[321]3/2$ at $\beta_2=0.5$ and $[330]1/2$ at $\beta_2=0.24$, show a strong soft dipole peaks near the threshold with the different peak heights: that  of the  $[330]1/2$ configuration is nearly 2 times larger  than that of the $[321]3/2$ configuration because of different  halo  $p$-wave amplitudes. This difference will help to assign the configuration, equivalently the magnitude of $\beta_2$ deformation if  the empirical $E1$ strength is  extracted from  the Coulomb breakup crosse sections, for example $^{31}$Ne$+^{208}$Pb reaction.

Another deformed halo nucleus $^{37}$Mg is also studied with the same deformed Woods-Saxon model.  The separation energy is constrained by recent empirical information of $S_n\sim200$ keV at a large deformation $\beta_2=0.46$. The last neutron configuration is assigned as $[321]1/2$ Nilsson orbit  which involves also large halo $p$-orbit configurations. The calculated $E1$ response shows a strong soft dipole nature.  
Compared with soft dipole strength of pure $p$-wave configuration in Fig . \ref{fig8}, 
the deformation effect quenches to some extent the peak of $E1$ strength distribution.
Experimental confirmation of soft dipole strength is highly desired to assign the deformation and the configuration of the last neutron orbit in $^{37}$Mg. 
Theoretically we will introduce more elaborate model such as deformed quai-particle random phase approximation (QRPA) on top of deformed Hartree-Fock model with continuum coupling or  
time-dependent finite amplitude approximation with DRHBc in near future \cite{PhysRevC.105.044312}.  

\begin{acknowledgments}
We thank Koichi Hagino for providing us with his deformed Woods-Saxon code. Fruitful discussions with Xiang-Xiang Sun, Yu-Ting Rong, Yi-Ming Jiang and Xin-Le Shang are gratefully acknowledged. 
This work is supported by the National Natural Science Foundation of China (Grant Nos. 12347139, 12047503, 12375118, 12435008, and W2412043), the Chinese Academy of Sciences (CAS) President’s International Fellowship Initiative (PIFI, Grant No. 2024PVA0003), the National Key R\&D Program of China (Grant Nos. 2023YFA1606503 and 2024YFE0109801) and the CAS Strategic Priority Research Program (Grant No. XDB34010000). The results described in this paper are obtained on the High-performance Computing Cluster of ITP-CAS and the ScGrid of the Supercomputing Center, Computer Network Information Center of Chinese Academy of Sciences. 
\end{acknowledgments}


\bibliography{apssamp.bib}

\begin{thebibliography}{62}%
\makeatletter
\providecommand \@ifxundefined [1]{%
 \@ifx{#1\undefined}
}%
\providecommand \@ifnum [1]{%
 \ifnum #1\expandafter \@firstoftwo
 \else \expandafter \@secondoftwo
 \fi
}%
\providecommand \@ifx [1]{%
 \ifx #1\expandafter \@firstoftwo
 \else \expandafter \@secondoftwo
 \fi
}%
\providecommand \natexlab [1]{#1}%
\providecommand \enquote  [1]{``#1''}%
\providecommand \bibnamefont  [1]{#1}%
\providecommand \bibfnamefont [1]{#1}%
\providecommand \citenamefont [1]{#1}%
\providecommand \href@noop [0]{\@secondoftwo}%
\providecommand \href [0]{\begingroup \@sanitize@url \@href}%
\providecommand \@href[1]{\@@startlink{#1}\@@href}%
\providecommand \@@href[1]{\endgroup#1\@@endlink}%
\providecommand \@sanitize@url [0]{\catcode `\\12\catcode `\$12\catcode `\&12\catcode `\#12\catcode `\^12\catcode `\_12\catcode `\%12\relax}%
\providecommand \@@startlink[1]{}%
\providecommand \@@endlink[0]{}%
\providecommand \url  [0]{\begingroup\@sanitize@url \@url }%
\providecommand \@url [1]{\endgroup\@href {#1}{\urlprefix }}%
\providecommand \urlprefix  [0]{URL }%
\providecommand \Eprint [0]{\href }%
\providecommand \doibase [0]{https://doi.org/}%
\providecommand \selectlanguage [0]{\@gobble}%
\providecommand \bibinfo  [0]{\@secondoftwo}%
\providecommand \bibfield  [0]{\@secondoftwo}%
\providecommand \translation [1]{[#1]}%
\providecommand \BibitemOpen [0]{}%
\providecommand \bibitemStop [0]{}%
\providecommand \bibitemNoStop [0]{.\EOS\space}%
\providecommand \EOS [0]{\spacefactor3000\relax}%
\providecommand \BibitemShut  [1]{\csname bibitem#1\endcsname}%
\let\auto@bib@innerbib\@empty
\bibitem [{\citenamefont {Tanihata}\ \emph {et~al.}(2013)\citenamefont {Tanihata}, \citenamefont {Savajols},\ and\ \citenamefont {Kanungo}}]{TANIHATA2013215}%
  \BibitemOpen
  \bibfield  {author} {\bibinfo {author} {\bibfnamefont {I.}~\bibnamefont {Tanihata}}, \bibinfo {author} {\bibfnamefont {H.}~\bibnamefont {Savajols}},\ and\ \bibinfo {author} {\bibfnamefont {R.}~\bibnamefont {Kanungo}},\ }\href {https://doi.org/https://doi.org/10.1016/j.ppnp.2012.07.001} {\bibfield  {journal} {\bibinfo  {journal} {Prog. Part. Nucl. Phys.}\ }\textbf {\bibinfo {volume} {68}},\ \bibinfo {pages} {215} (\bibinfo {year} {2013})}\BibitemShut {NoStop}%
\bibitem [{\citenamefont {Zhou}(2017)}]{Zhou:2017ldu}%
  \BibitemOpen
  \bibfield  {author} {\bibinfo {author} {\bibfnamefont {S.-G.}\ \bibnamefont {Zhou}},\ }\href {https://doi.org/10.22323/1.281.0373} {\bibfield  {journal} {\bibinfo  {journal} {PoS}\ }\textbf {\bibinfo {volume} {INPC2016}},\ \bibinfo {pages} {373} (\bibinfo {year} {2017})},\ \Eprint {https://arxiv.org/abs/1703.09045} {arXiv:1703.09045 [nucl-th]} \BibitemShut {NoStop}%
\bibitem [{\citenamefont {Ye}\ \emph {et~al.}(2025)\citenamefont {Ye}, \citenamefont {Yang}, \citenamefont {Sakurai},\ and\ \citenamefont {Hu}}]{ye24}%
  \BibitemOpen
  \bibfield  {author} {\bibinfo {author} {\bibfnamefont {Y.}~\bibnamefont {Ye}}, \bibinfo {author} {\bibfnamefont {X.}~\bibnamefont {Yang}}, \bibinfo {author} {\bibfnamefont {H.}~\bibnamefont {Sakurai}},\ and\ \bibinfo {author} {\bibfnamefont {B.}~\bibnamefont {Hu}},\ }\href {https://doi.org/10.1038/s42254-024-00782-5} {\bibfield  {journal} {\bibinfo  {journal} {Nat. Rev. Phys}\ }\textbf {\bibinfo {volume} {7}},\ \bibinfo {pages} {21} (\bibinfo {year} {2025})}\BibitemShut {NoStop}%
\bibitem [{\citenamefont {Tanihata}\ \emph {et~al.}(1985)\citenamefont {Tanihata}, \citenamefont {Hamagaki}, \citenamefont {Hashimoto}, \citenamefont {Shida}, \citenamefont {Yoshikawa}, \citenamefont {Sugimoto}, \citenamefont {Yamakawa}, \citenamefont {Kobayashi},\ and\ \citenamefont {Takahashi}}]{Tanihata1985}%
  \BibitemOpen
  \bibfield  {author} {\bibinfo {author} {\bibfnamefont {I.}~\bibnamefont {Tanihata}}, \bibinfo {author} {\bibfnamefont {H.}~\bibnamefont {Hamagaki}}, \bibinfo {author} {\bibfnamefont {O.}~\bibnamefont {Hashimoto}}, \bibinfo {author} {\bibfnamefont {Y.}~\bibnamefont {Shida}}, \bibinfo {author} {\bibfnamefont {N.}~\bibnamefont {Yoshikawa}}, \bibinfo {author} {\bibfnamefont {K.}~\bibnamefont {Sugimoto}}, \bibinfo {author} {\bibfnamefont {O.}~\bibnamefont {Yamakawa}}, \bibinfo {author} {\bibfnamefont {T.}~\bibnamefont {Kobayashi}},\ and\ \bibinfo {author} {\bibfnamefont {N.}~\bibnamefont {Takahashi}},\ }\href {https://doi.org/10.1103/PhysRevLett.55.2676} {\bibfield  {journal} {\bibinfo  {journal} {Phys. Rev. Lett.}\ }\textbf {\bibinfo {volume} {55}},\ \bibinfo {pages} {2676} (\bibinfo {year} {1985})}\BibitemShut {NoStop}%
\bibitem [{\citenamefont {Tanihata}\ \emph {et~al.}(1988)\citenamefont {Tanihata}, \citenamefont {Kobayashi}, \citenamefont {Yamakawa}, \citenamefont {Shimoura}, \citenamefont {Ekuni}, \citenamefont {Sugimoto}, \citenamefont {Takahashi}, \citenamefont {Shimoda},\ and\ \citenamefont {Sato}}]{TANIHATA1988}%
  \BibitemOpen
  \bibfield  {author} {\bibinfo {author} {\bibfnamefont {I.}~\bibnamefont {Tanihata}}, \bibinfo {author} {\bibfnamefont {T.}~\bibnamefont {Kobayashi}}, \bibinfo {author} {\bibfnamefont {O.}~\bibnamefont {Yamakawa}}, \bibinfo {author} {\bibfnamefont {S.}~\bibnamefont {Shimoura}}, \bibinfo {author} {\bibfnamefont {K.}~\bibnamefont {Ekuni}}, \bibinfo {author} {\bibfnamefont {K.}~\bibnamefont {Sugimoto}}, \bibinfo {author} {\bibfnamefont {N.}~\bibnamefont {Takahashi}}, \bibinfo {author} {\bibfnamefont {T.}~\bibnamefont {Shimoda}},\ and\ \bibinfo {author} {\bibfnamefont {H.}~\bibnamefont {Sato}},\ }\href {https://doi.org/https://doi.org/10.1016/0370-2693(88)90702-2} {\bibfield  {journal} {\bibinfo  {journal} {Phys. Lett. B}\ }\textbf {\bibinfo {volume} {206}},\ \bibinfo {pages} {592} (\bibinfo {year} {1988})}\BibitemShut {NoStop}%
\bibitem [{\citenamefont {Fukuda}\ \emph {et~al.}(1991)\citenamefont {Fukuda}, \citenamefont {Ichihara}, \citenamefont {Inabe}, \citenamefont {Kubo}, \citenamefont {Kumagai}, \citenamefont {Nakagawa}, \citenamefont {Yano}, \citenamefont {Tanihata}, \citenamefont {Adachi}, \citenamefont {Asahi}, \citenamefont {Kouguchi}, \citenamefont {Ishihara}, \citenamefont {Sagawa},\ and\ \citenamefont {Shimoura}}]{FUKUDA1991}%
  \BibitemOpen
  \bibfield  {author} {\bibinfo {author} {\bibfnamefont {M.}~\bibnamefont {Fukuda}}, \bibinfo {author} {\bibfnamefont {T.}~\bibnamefont {Ichihara}}, \bibinfo {author} {\bibfnamefont {N.}~\bibnamefont {Inabe}}, \bibinfo {author} {\bibfnamefont {T.}~\bibnamefont {Kubo}}, \bibinfo {author} {\bibfnamefont {H.}~\bibnamefont {Kumagai}}, \bibinfo {author} {\bibfnamefont {T.}~\bibnamefont {Nakagawa}}, \bibinfo {author} {\bibfnamefont {Y.}~\bibnamefont {Yano}}, \bibinfo {author} {\bibfnamefont {I.}~\bibnamefont {Tanihata}}, \bibinfo {author} {\bibfnamefont {M.}~\bibnamefont {Adachi}}, \bibinfo {author} {\bibfnamefont {K.}~\bibnamefont {Asahi}}, \bibinfo {author} {\bibfnamefont {M.}~\bibnamefont {Kouguchi}}, \bibinfo {author} {\bibfnamefont {M.}~\bibnamefont {Ishihara}}, \bibinfo {author} {\bibfnamefont {H.}~\bibnamefont {Sagawa}},\ and\ \bibinfo {author} {\bibfnamefont {S.}~\bibnamefont {Shimoura}},\ }\href {https://doi.org/https://doi.org/10.1016/0370-2693(91)91587-L} {\bibfield  {journal} {\bibinfo  {journal} {Phys.
  Lett. B}\ }\textbf {\bibinfo {volume} {268}},\ \bibinfo {pages} {339} (\bibinfo {year} {1991})}\BibitemShut {NoStop}%
\bibitem [{\citenamefont {Bazin}\ \emph {et~al.}(1995)\citenamefont {Bazin}, \citenamefont {Brown}, \citenamefont {Brown}, \citenamefont {Fauerbach}, \citenamefont {Hellström}, \citenamefont {Hirzebruch}, \citenamefont {Kelley}, \citenamefont {Kryger}, \citenamefont {Morrissey}, \citenamefont {Pfaff}, \citenamefont {Powell}, \citenamefont {Sherrill},\ and\ \citenamefont {Thoennessen}}]{19c1995}%
  \BibitemOpen
  \bibfield  {author} {\bibinfo {author} {\bibfnamefont {D.}~\bibnamefont {Bazin}}, \bibinfo {author} {\bibfnamefont {B.}~\bibnamefont {Brown}}, \bibinfo {author} {\bibfnamefont {J.}~\bibnamefont {Brown}}, \bibinfo {author} {\bibfnamefont {M.}~\bibnamefont {Fauerbach}}, \bibinfo {author} {\bibfnamefont {M.}~\bibnamefont {Hellström}}, \bibinfo {author} {\bibfnamefont {S.}~\bibnamefont {Hirzebruch}}, \bibinfo {author} {\bibfnamefont {J.}~\bibnamefont {Kelley}}, \bibinfo {author} {\bibfnamefont {R.}~\bibnamefont {Kryger}}, \bibinfo {author} {\bibfnamefont {D.}~\bibnamefont {Morrissey}}, \bibinfo {author} {\bibfnamefont {R.}~\bibnamefont {Pfaff}}, \bibinfo {author} {\bibfnamefont {C.}~\bibnamefont {Powell}}, \bibinfo {author} {\bibfnamefont {B.}~\bibnamefont {Sherrill}},\ and\ \bibinfo {author} {\bibfnamefont {M.}~\bibnamefont {Thoennessen}},\ }\href {https://doi.org/10.1103/PhysRevLett.74.3569} {\bibfield  {journal} {\bibinfo  {journal} {Phys. Rev. Lett.}\ }\textbf {\bibinfo {volume} {74}},\ \bibinfo {pages}
  {3569} (\bibinfo {year} {1995})}\BibitemShut {NoStop}%
\bibitem [{\citenamefont {Nakamura}\ \emph {et~al.}(1999)\citenamefont {Nakamura}, \citenamefont {Fukuda}, \citenamefont {Kobayashi}, \citenamefont {Aoi}, \citenamefont {Iwasaki}, \citenamefont {Kubo}, \citenamefont {Mengoni}, \citenamefont {Notani}, \citenamefont {Otsu}, \citenamefont {Sakurai}, \citenamefont {Shimoura}, \citenamefont {Teranishi}, \citenamefont {Watanabe}, \citenamefont {Yoneda},\ and\ \citenamefont {Ishihara}}]{Nakamura1999}%
  \BibitemOpen
  \bibfield  {author} {\bibinfo {author} {\bibfnamefont {T.}~\bibnamefont {Nakamura}}, \bibinfo {author} {\bibfnamefont {N.}~\bibnamefont {Fukuda}}, \bibinfo {author} {\bibfnamefont {T.}~\bibnamefont {Kobayashi}}, \bibinfo {author} {\bibfnamefont {N.}~\bibnamefont {Aoi}}, \bibinfo {author} {\bibfnamefont {H.}~\bibnamefont {Iwasaki}}, \bibinfo {author} {\bibfnamefont {T.}~\bibnamefont {Kubo}}, \bibinfo {author} {\bibfnamefont {A.}~\bibnamefont {Mengoni}}, \bibinfo {author} {\bibfnamefont {M.}~\bibnamefont {Notani}}, \bibinfo {author} {\bibfnamefont {H.}~\bibnamefont {Otsu}}, \bibinfo {author} {\bibfnamefont {H.}~\bibnamefont {Sakurai}}, \bibinfo {author} {\bibfnamefont {S.}~\bibnamefont {Shimoura}}, \bibinfo {author} {\bibfnamefont {T.}~\bibnamefont {Teranishi}}, \bibinfo {author} {\bibfnamefont {Y.~X.}\ \bibnamefont {Watanabe}}, \bibinfo {author} {\bibfnamefont {K.}~\bibnamefont {Yoneda}},\ and\ \bibinfo {author} {\bibfnamefont {M.}~\bibnamefont {Ishihara}},\ }\href
  {https://doi.org/10.1103/PhysRevLett.83.1112} {\bibfield  {journal} {\bibinfo  {journal} {Phys. Rev. Lett.}\ }\textbf {\bibinfo {volume} {83}},\ \bibinfo {pages} {1112} (\bibinfo {year} {1999})}\BibitemShut {NoStop}%
\bibitem [{\citenamefont {Nakamura}\ \emph {et~al.}(2009)\citenamefont {Nakamura}, \citenamefont {Kobayashi}, \citenamefont {Kondo}, \citenamefont {Satou}, \citenamefont {Aoi}, \citenamefont {Baba}, \citenamefont {Deguchi}, \citenamefont {Fukuda}, \citenamefont {Gibelin}, \citenamefont {Inabe}, \citenamefont {Ishihara}, \citenamefont {Kameda}, \citenamefont {Kawada}, \citenamefont {Kubo}, \citenamefont {Kusaka}, \citenamefont {Mengoni}, \citenamefont {Motobayashi}, \citenamefont {Ohnishi}, \citenamefont {Ohtake}, \citenamefont {Orr}, \citenamefont {Otsu}, \citenamefont {Otsuka}, \citenamefont {Saito}, \citenamefont {Sakurai}, \citenamefont {Shimoura}, \citenamefont {Sumikama}, \citenamefont {Takeda}, \citenamefont {Takeshita}, \citenamefont {Takechi}, \citenamefont {Takeuchi}, \citenamefont {Tanaka}, \citenamefont {Tanaka}, \citenamefont {Tanaka}, \citenamefont {Togano}, \citenamefont {Utsuno}, \citenamefont {Yoneda}, \citenamefont {Yoshida},\ and\ \citenamefont {Yoshida}}]{Nakamura2009}%
  \BibitemOpen
  \bibfield  {author} {\bibinfo {author} {\bibfnamefont {T.}~\bibnamefont {Nakamura}}, \bibinfo {author} {\bibfnamefont {N.}~\bibnamefont {Kobayashi}}, \bibinfo {author} {\bibfnamefont {Y.}~\bibnamefont {Kondo}}, \bibinfo {author} {\bibfnamefont {Y.}~\bibnamefont {Satou}}, \bibinfo {author} {\bibfnamefont {N.}~\bibnamefont {Aoi}}, \bibinfo {author} {\bibfnamefont {H.}~\bibnamefont {Baba}}, \bibinfo {author} {\bibfnamefont {S.}~\bibnamefont {Deguchi}}, \bibinfo {author} {\bibfnamefont {N.}~\bibnamefont {Fukuda}}, \bibinfo {author} {\bibfnamefont {J.}~\bibnamefont {Gibelin}}, \bibinfo {author} {\bibfnamefont {N.}~\bibnamefont {Inabe}}, \bibinfo {author} {\bibfnamefont {M.}~\bibnamefont {Ishihara}}, \bibinfo {author} {\bibfnamefont {D.}~\bibnamefont {Kameda}}, \bibinfo {author} {\bibfnamefont {Y.}~\bibnamefont {Kawada}}, \bibinfo {author} {\bibfnamefont {T.}~\bibnamefont {Kubo}}, \bibinfo {author} {\bibfnamefont {K.}~\bibnamefont {Kusaka}}, \bibinfo {author} {\bibfnamefont {A.}~\bibnamefont {Mengoni}}, \bibinfo
  {author} {\bibfnamefont {T.}~\bibnamefont {Motobayashi}}, \bibinfo {author} {\bibfnamefont {T.}~\bibnamefont {Ohnishi}}, \bibinfo {author} {\bibfnamefont {M.}~\bibnamefont {Ohtake}}, \bibinfo {author} {\bibfnamefont {N.~A.}\ \bibnamefont {Orr}}, \bibinfo {author} {\bibfnamefont {H.}~\bibnamefont {Otsu}}, \bibinfo {author} {\bibfnamefont {T.}~\bibnamefont {Otsuka}}, \bibinfo {author} {\bibfnamefont {A.}~\bibnamefont {Saito}}, \bibinfo {author} {\bibfnamefont {H.}~\bibnamefont {Sakurai}}, \bibinfo {author} {\bibfnamefont {S.}~\bibnamefont {Shimoura}}, \bibinfo {author} {\bibfnamefont {T.}~\bibnamefont {Sumikama}}, \bibinfo {author} {\bibfnamefont {H.}~\bibnamefont {Takeda}}, \bibinfo {author} {\bibfnamefont {E.}~\bibnamefont {Takeshita}}, \bibinfo {author} {\bibfnamefont {M.}~\bibnamefont {Takechi}}, \bibinfo {author} {\bibfnamefont {S.}~\bibnamefont {Takeuchi}}, \bibinfo {author} {\bibfnamefont {K.}~\bibnamefont {Tanaka}}, \bibinfo {author} {\bibfnamefont {K.~N.}\ \bibnamefont {Tanaka}}, \bibinfo {author}
  {\bibfnamefont {N.}~\bibnamefont {Tanaka}}, \bibinfo {author} {\bibfnamefont {Y.}~\bibnamefont {Togano}}, \bibinfo {author} {\bibfnamefont {Y.}~\bibnamefont {Utsuno}}, \bibinfo {author} {\bibfnamefont {K.}~\bibnamefont {Yoneda}}, \bibinfo {author} {\bibfnamefont {A.}~\bibnamefont {Yoshida}},\ and\ \bibinfo {author} {\bibfnamefont {K.}~\bibnamefont {Yoshida}},\ }\href {https://doi.org/10.1103/PhysRevLett.103.262501} {\bibfield  {journal} {\bibinfo  {journal} {Phys. Rev. Lett.}\ }\textbf {\bibinfo {volume} {103}},\ \bibinfo {pages} {262501} (\bibinfo {year} {2009})}\BibitemShut {NoStop}%
\bibitem [{\citenamefont {Nakamura}\ \emph {et~al.}(2014)\citenamefont {Nakamura}, \citenamefont {Kobayashi}, \citenamefont {Kondo}, \citenamefont {Satou}, \citenamefont {Tostevin}, \citenamefont {Utsuno}, \citenamefont {Aoi}, \citenamefont {Baba}, \citenamefont {Fukuda}, \citenamefont {Gibelin}, \citenamefont {Inabe}, \citenamefont {Ishihara}, \citenamefont {Kameda}, \citenamefont {Kubo}, \citenamefont {Motobayashi}, \citenamefont {Ohnishi}, \citenamefont {Orr}, \citenamefont {Otsu}, \citenamefont {Otsuka}, \citenamefont {Sakurai}, \citenamefont {Sumikama}, \citenamefont {Takeda}, \citenamefont {Takeshita}, \citenamefont {Takechi}, \citenamefont {Takeuchi}, \citenamefont {Togano},\ and\ \citenamefont {Yoneda}}]{Nakamura2014}%
  \BibitemOpen
  \bibfield  {author} {\bibinfo {author} {\bibfnamefont {T.}~\bibnamefont {Nakamura}}, \bibinfo {author} {\bibfnamefont {N.}~\bibnamefont {Kobayashi}}, \bibinfo {author} {\bibfnamefont {Y.}~\bibnamefont {Kondo}}, \bibinfo {author} {\bibfnamefont {Y.}~\bibnamefont {Satou}}, \bibinfo {author} {\bibfnamefont {J.~A.}\ \bibnamefont {Tostevin}}, \bibinfo {author} {\bibfnamefont {Y.}~\bibnamefont {Utsuno}}, \bibinfo {author} {\bibfnamefont {N.}~\bibnamefont {Aoi}}, \bibinfo {author} {\bibfnamefont {H.}~\bibnamefont {Baba}}, \bibinfo {author} {\bibfnamefont {N.}~\bibnamefont {Fukuda}}, \bibinfo {author} {\bibfnamefont {J.}~\bibnamefont {Gibelin}}, \bibinfo {author} {\bibfnamefont {N.}~\bibnamefont {Inabe}}, \bibinfo {author} {\bibfnamefont {M.}~\bibnamefont {Ishihara}}, \bibinfo {author} {\bibfnamefont {D.}~\bibnamefont {Kameda}}, \bibinfo {author} {\bibfnamefont {T.}~\bibnamefont {Kubo}}, \bibinfo {author} {\bibfnamefont {T.}~\bibnamefont {Motobayashi}}, \bibinfo {author} {\bibfnamefont {T.}~\bibnamefont {Ohnishi}},
  \bibinfo {author} {\bibfnamefont {N.~A.}\ \bibnamefont {Orr}}, \bibinfo {author} {\bibfnamefont {H.}~\bibnamefont {Otsu}}, \bibinfo {author} {\bibfnamefont {T.}~\bibnamefont {Otsuka}}, \bibinfo {author} {\bibfnamefont {H.}~\bibnamefont {Sakurai}}, \bibinfo {author} {\bibfnamefont {T.}~\bibnamefont {Sumikama}}, \bibinfo {author} {\bibfnamefont {H.}~\bibnamefont {Takeda}}, \bibinfo {author} {\bibfnamefont {E.}~\bibnamefont {Takeshita}}, \bibinfo {author} {\bibfnamefont {M.}~\bibnamefont {Takechi}}, \bibinfo {author} {\bibfnamefont {S.}~\bibnamefont {Takeuchi}}, \bibinfo {author} {\bibfnamefont {Y.}~\bibnamefont {Togano}},\ and\ \bibinfo {author} {\bibfnamefont {K.}~\bibnamefont {Yoneda}},\ }\href {https://doi.org/10.1103/PhysRevLett.112.142501} {\bibfield  {journal} {\bibinfo  {journal} {Phys. Rev. Lett.}\ }\textbf {\bibinfo {volume} {112}},\ \bibinfo {pages} {142501} (\bibinfo {year} {2014})}\BibitemShut {NoStop}%
\bibitem [{\citenamefont {Kobayashi}\ \emph {et~al.}(2014)\citenamefont {Kobayashi}, \citenamefont {Nakamura}, \citenamefont {Kondo}, \citenamefont {Tostevin}, \citenamefont {Utsuno}, \citenamefont {Aoi}, \citenamefont {Baba}, \citenamefont {Barthelemy}, \citenamefont {Famiano}, \citenamefont {Fukuda}, \citenamefont {Inabe}, \citenamefont {Ishihara}, \citenamefont {Kanungo}, \citenamefont {Kim}, \citenamefont {Kubo}, \citenamefont {Lee}, \citenamefont {Lee}, \citenamefont {Matsushita}, \citenamefont {Motobayashi}, \citenamefont {Ohnishi}, \citenamefont {Orr}, \citenamefont {Otsu}, \citenamefont {Otsuka}, \citenamefont {Sako}, \citenamefont {Sakurai}, \citenamefont {Satou}, \citenamefont {Sumikama}, \citenamefont {Takeda}, \citenamefont {Takeuchi}, \citenamefont {Tanaka}, \citenamefont {Togano},\ and\ \citenamefont {Yoneda}}]{mg37}%
  \BibitemOpen
  \bibfield  {author} {\bibinfo {author} {\bibfnamefont {N.}~\bibnamefont {Kobayashi}}, \bibinfo {author} {\bibfnamefont {T.}~\bibnamefont {Nakamura}}, \bibinfo {author} {\bibfnamefont {Y.}~\bibnamefont {Kondo}}, \bibinfo {author} {\bibfnamefont {J.~A.}\ \bibnamefont {Tostevin}}, \bibinfo {author} {\bibfnamefont {Y.}~\bibnamefont {Utsuno}}, \bibinfo {author} {\bibfnamefont {N.}~\bibnamefont {Aoi}}, \bibinfo {author} {\bibfnamefont {H.}~\bibnamefont {Baba}}, \bibinfo {author} {\bibfnamefont {R.}~\bibnamefont {Barthelemy}}, \bibinfo {author} {\bibfnamefont {M.~A.}\ \bibnamefont {Famiano}}, \bibinfo {author} {\bibfnamefont {N.}~\bibnamefont {Fukuda}}, \bibinfo {author} {\bibfnamefont {N.}~\bibnamefont {Inabe}}, \bibinfo {author} {\bibfnamefont {M.}~\bibnamefont {Ishihara}}, \bibinfo {author} {\bibfnamefont {R.}~\bibnamefont {Kanungo}}, \bibinfo {author} {\bibfnamefont {S.}~\bibnamefont {Kim}}, \bibinfo {author} {\bibfnamefont {T.}~\bibnamefont {Kubo}}, \bibinfo {author} {\bibfnamefont {G.~S.}\ \bibnamefont
  {Lee}}, \bibinfo {author} {\bibfnamefont {H.~S.}\ \bibnamefont {Lee}}, \bibinfo {author} {\bibfnamefont {M.}~\bibnamefont {Matsushita}}, \bibinfo {author} {\bibfnamefont {T.}~\bibnamefont {Motobayashi}}, \bibinfo {author} {\bibfnamefont {T.}~\bibnamefont {Ohnishi}}, \bibinfo {author} {\bibfnamefont {N.~A.}\ \bibnamefont {Orr}}, \bibinfo {author} {\bibfnamefont {H.}~\bibnamefont {Otsu}}, \bibinfo {author} {\bibfnamefont {T.}~\bibnamefont {Otsuka}}, \bibinfo {author} {\bibfnamefont {T.}~\bibnamefont {Sako}}, \bibinfo {author} {\bibfnamefont {H.}~\bibnamefont {Sakurai}}, \bibinfo {author} {\bibfnamefont {Y.}~\bibnamefont {Satou}}, \bibinfo {author} {\bibfnamefont {T.}~\bibnamefont {Sumikama}}, \bibinfo {author} {\bibfnamefont {H.}~\bibnamefont {Takeda}}, \bibinfo {author} {\bibfnamefont {S.}~\bibnamefont {Takeuchi}}, \bibinfo {author} {\bibfnamefont {R.}~\bibnamefont {Tanaka}}, \bibinfo {author} {\bibfnamefont {Y.}~\bibnamefont {Togano}},\ and\ \bibinfo {author} {\bibfnamefont {K.}~\bibnamefont {Yoneda}},\
  }\href {https://doi.org/10.1103/PhysRevLett.112.242501} {\bibfield  {journal} {\bibinfo  {journal} {Phys. Rev. Lett.}\ }\textbf {\bibinfo {volume} {112}},\ \bibinfo {pages} {242501} (\bibinfo {year} {2014})}\BibitemShut {NoStop}%
\bibitem [{\citenamefont {Takechi}\ \emph {et~al.}(2014)\citenamefont {Takechi}, \citenamefont {Suzuki}, \citenamefont {Nishimura}, \citenamefont {Fukuda}, \citenamefont {Ohtsubo}, \citenamefont {Nagashima}, \citenamefont {Suzuki}, \citenamefont {Yamaguchi}, \citenamefont {Ozawa}, \citenamefont {Moriguchi}, \citenamefont {Ohishi}, \citenamefont {Sumikama}, \citenamefont {Geissel}, \citenamefont {Aoi}, \citenamefont {Chen}, \citenamefont {Fang}, \citenamefont {Fukuda}, \citenamefont {Fukuoka}, \citenamefont {Furuki}, \citenamefont {Inabe}, \citenamefont {Ishibashi}, \citenamefont {Itoh}, \citenamefont {Izumikawa}, \citenamefont {Kameda}, \citenamefont {Kubo}, \citenamefont {Lantz}, \citenamefont {Lee}, \citenamefont {Ma}, \citenamefont {Matsuta}, \citenamefont {Mihara}, \citenamefont {Momota}, \citenamefont {Nagae}, \citenamefont {Nishikiori}, \citenamefont {Niwa}, \citenamefont {Ohnishi}, \citenamefont {Okumura}, \citenamefont {Ohtake}, \citenamefont {Ogura}, \citenamefont {Sakurai}, \citenamefont {Sato},
  \citenamefont {Shimbara}, \citenamefont {Suzuki}, \citenamefont {Takeda}, \citenamefont {Takeuchi}, \citenamefont {Tanaka}, \citenamefont {Tanaka}, \citenamefont {Uenishi}, \citenamefont {Winkler}, \citenamefont {Yanagisawa}, \citenamefont {Watanabe}, \citenamefont {Minomo}, \citenamefont {Tagami}, \citenamefont {Shimada}, \citenamefont {Kimura}, \citenamefont {Matsumoto}, \citenamefont {Shimizu},\ and\ \citenamefont {Yahiro}}]{mg371}%
  \BibitemOpen
  \bibfield  {author} {\bibinfo {author} {\bibfnamefont {M.}~\bibnamefont {Takechi}}, \bibinfo {author} {\bibfnamefont {S.}~\bibnamefont {Suzuki}}, \bibinfo {author} {\bibfnamefont {D.}~\bibnamefont {Nishimura}}, \bibinfo {author} {\bibfnamefont {M.}~\bibnamefont {Fukuda}}, \bibinfo {author} {\bibfnamefont {T.}~\bibnamefont {Ohtsubo}}, \bibinfo {author} {\bibfnamefont {M.}~\bibnamefont {Nagashima}}, \bibinfo {author} {\bibfnamefont {T.}~\bibnamefont {Suzuki}}, \bibinfo {author} {\bibfnamefont {T.}~\bibnamefont {Yamaguchi}}, \bibinfo {author} {\bibfnamefont {A.}~\bibnamefont {Ozawa}}, \bibinfo {author} {\bibfnamefont {T.}~\bibnamefont {Moriguchi}}, \bibinfo {author} {\bibfnamefont {H.}~\bibnamefont {Ohishi}}, \bibinfo {author} {\bibfnamefont {T.}~\bibnamefont {Sumikama}}, \bibinfo {author} {\bibfnamefont {H.}~\bibnamefont {Geissel}}, \bibinfo {author} {\bibfnamefont {N.}~\bibnamefont {Aoi}}, \bibinfo {author} {\bibfnamefont {R.-J.}\ \bibnamefont {Chen}}, \bibinfo {author} {\bibfnamefont {D.-Q.}\ \bibnamefont
  {Fang}}, \bibinfo {author} {\bibfnamefont {N.}~\bibnamefont {Fukuda}}, \bibinfo {author} {\bibfnamefont {S.}~\bibnamefont {Fukuoka}}, \bibinfo {author} {\bibfnamefont {H.}~\bibnamefont {Furuki}}, \bibinfo {author} {\bibfnamefont {N.}~\bibnamefont {Inabe}}, \bibinfo {author} {\bibfnamefont {Y.}~\bibnamefont {Ishibashi}}, \bibinfo {author} {\bibfnamefont {T.}~\bibnamefont {Itoh}}, \bibinfo {author} {\bibfnamefont {T.}~\bibnamefont {Izumikawa}}, \bibinfo {author} {\bibfnamefont {D.}~\bibnamefont {Kameda}}, \bibinfo {author} {\bibfnamefont {T.}~\bibnamefont {Kubo}}, \bibinfo {author} {\bibfnamefont {M.}~\bibnamefont {Lantz}}, \bibinfo {author} {\bibfnamefont {C.~S.}\ \bibnamefont {Lee}}, \bibinfo {author} {\bibfnamefont {Y.-G.}\ \bibnamefont {Ma}}, \bibinfo {author} {\bibfnamefont {K.}~\bibnamefont {Matsuta}}, \bibinfo {author} {\bibfnamefont {M.}~\bibnamefont {Mihara}}, \bibinfo {author} {\bibfnamefont {S.}~\bibnamefont {Momota}}, \bibinfo {author} {\bibfnamefont {D.}~\bibnamefont {Nagae}}, \bibinfo {author}
  {\bibfnamefont {R.}~\bibnamefont {Nishikiori}}, \bibinfo {author} {\bibfnamefont {T.}~\bibnamefont {Niwa}}, \bibinfo {author} {\bibfnamefont {T.}~\bibnamefont {Ohnishi}}, \bibinfo {author} {\bibfnamefont {K.}~\bibnamefont {Okumura}}, \bibinfo {author} {\bibfnamefont {M.}~\bibnamefont {Ohtake}}, \bibinfo {author} {\bibfnamefont {T.}~\bibnamefont {Ogura}}, \bibinfo {author} {\bibfnamefont {H.}~\bibnamefont {Sakurai}}, \bibinfo {author} {\bibfnamefont {K.}~\bibnamefont {Sato}}, \bibinfo {author} {\bibfnamefont {Y.}~\bibnamefont {Shimbara}}, \bibinfo {author} {\bibfnamefont {H.}~\bibnamefont {Suzuki}}, \bibinfo {author} {\bibfnamefont {H.}~\bibnamefont {Takeda}}, \bibinfo {author} {\bibfnamefont {S.}~\bibnamefont {Takeuchi}}, \bibinfo {author} {\bibfnamefont {K.}~\bibnamefont {Tanaka}}, \bibinfo {author} {\bibfnamefont {M.}~\bibnamefont {Tanaka}}, \bibinfo {author} {\bibfnamefont {H.}~\bibnamefont {Uenishi}}, \bibinfo {author} {\bibfnamefont {M.}~\bibnamefont {Winkler}}, \bibinfo {author} {\bibfnamefont
  {Y.}~\bibnamefont {Yanagisawa}}, \bibinfo {author} {\bibfnamefont {S.}~\bibnamefont {Watanabe}}, \bibinfo {author} {\bibfnamefont {K.}~\bibnamefont {Minomo}}, \bibinfo {author} {\bibfnamefont {S.}~\bibnamefont {Tagami}}, \bibinfo {author} {\bibfnamefont {M.}~\bibnamefont {Shimada}}, \bibinfo {author} {\bibfnamefont {M.}~\bibnamefont {Kimura}}, \bibinfo {author} {\bibfnamefont {T.}~\bibnamefont {Matsumoto}}, \bibinfo {author} {\bibfnamefont {Y.~R.}\ \bibnamefont {Shimizu}},\ and\ \bibinfo {author} {\bibfnamefont {M.}~\bibnamefont {Yahiro}},\ }\href {https://doi.org/10.1103/PhysRevC.90.061305} {\bibfield  {journal} {\bibinfo  {journal} {Phys. Rev. C}\ }\textbf {\bibinfo {volume} {90}},\ \bibinfo {pages} {061305} (\bibinfo {year} {2014})}\BibitemShut {NoStop}%
\bibitem [{\citenamefont {Suzuki}\ \emph {et~al.}(2002)\citenamefont {Suzuki}, \citenamefont {Ogawa}, \citenamefont {Chiba}, \citenamefont {Fukuda}, \citenamefont {Iwasa}, \citenamefont {Izumikawa}, \citenamefont {Kanungo}, \citenamefont {Kawamura}, \citenamefont {Ozawa}, \citenamefont {Suda}, \citenamefont {Tanihata}, \citenamefont {Watanabe}, \citenamefont {Yamaguchi},\ and\ \citenamefont {Yamaguchi}}]{17B02}%
  \BibitemOpen
  \bibfield  {author} {\bibinfo {author} {\bibfnamefont {T.}~\bibnamefont {Suzuki}}, \bibinfo {author} {\bibfnamefont {Y.}~\bibnamefont {Ogawa}}, \bibinfo {author} {\bibfnamefont {M.}~\bibnamefont {Chiba}}, \bibinfo {author} {\bibfnamefont {M.}~\bibnamefont {Fukuda}}, \bibinfo {author} {\bibfnamefont {N.}~\bibnamefont {Iwasa}}, \bibinfo {author} {\bibfnamefont {T.}~\bibnamefont {Izumikawa}}, \bibinfo {author} {\bibfnamefont {R.}~\bibnamefont {Kanungo}}, \bibinfo {author} {\bibfnamefont {Y.}~\bibnamefont {Kawamura}}, \bibinfo {author} {\bibfnamefont {A.}~\bibnamefont {Ozawa}}, \bibinfo {author} {\bibfnamefont {T.}~\bibnamefont {Suda}}, \bibinfo {author} {\bibfnamefont {I.}~\bibnamefont {Tanihata}}, \bibinfo {author} {\bibfnamefont {S.}~\bibnamefont {Watanabe}}, \bibinfo {author} {\bibfnamefont {T.}~\bibnamefont {Yamaguchi}},\ and\ \bibinfo {author} {\bibfnamefont {Y.}~\bibnamefont {Yamaguchi}},\ }\href {https://doi.org/10.1103/PhysRevLett.89.012501} {\bibfield  {journal} {\bibinfo  {journal} {Phys. Rev.
  Lett.}\ }\textbf {\bibinfo {volume} {89}},\ \bibinfo {pages} {012501} (\bibinfo {year} {2002})}\BibitemShut {NoStop}%
\bibitem [{\citenamefont {Yang}\ \emph {et~al.}(2021)\citenamefont {Yang}, \citenamefont {Kubota}, \citenamefont {Corsi}, \citenamefont {Yoshida}, \citenamefont {Sun}, \citenamefont {Li}, \citenamefont {Kimura}, \citenamefont {Michel}, \citenamefont {Ogata}, \citenamefont {Yuan}, \citenamefont {Yuan}, \citenamefont {Authelet}, \citenamefont {Baba}, \citenamefont {Caesar}, \citenamefont {Calvet}, \citenamefont {Delbart}, \citenamefont {Dozono}, \citenamefont {Feng}, \citenamefont {Flavigny}, \citenamefont {Gheller}, \citenamefont {Gibelin}, \citenamefont {Giganon}, \citenamefont {Gillibert}, \citenamefont {Hasegawa}, \citenamefont {Isobe}, \citenamefont {Kanaya}, \citenamefont {Kawakami}, \citenamefont {Kim}, \citenamefont {Kiyokawa}, \citenamefont {Kobayashi}, \citenamefont {Kobayashi}, \citenamefont {Kobayashi}, \citenamefont {Kondo}, \citenamefont {Korkulu}, \citenamefont {Koyama}, \citenamefont {Lapoux}, \citenamefont {Maeda}, \citenamefont {Marqu\'es}, \citenamefont {Motobayashi}, \citenamefont
  {Miyazaki}, \citenamefont {Nakamura}, \citenamefont {Nakatsuka}, \citenamefont {Nishio}, \citenamefont {Obertelli}, \citenamefont {Ohkura}, \citenamefont {Orr}, \citenamefont {Ota}, \citenamefont {Otsu}, \citenamefont {Ozaki}, \citenamefont {Panin}, \citenamefont {Paschalis}, \citenamefont {Pollacco}, \citenamefont {Reichert}, \citenamefont {Rouss\'e}, \citenamefont {Saito}, \citenamefont {Sakaguchi}, \citenamefont {Sako}, \citenamefont {Santamaria}, \citenamefont {Sasano}, \citenamefont {Sato}, \citenamefont {Shikata}, \citenamefont {Shimizu}, \citenamefont {Shindo}, \citenamefont {Stuhl}, \citenamefont {Sumikama}, \citenamefont {Sun}, \citenamefont {Tabata}, \citenamefont {Togano}, \citenamefont {Tsubota}, \citenamefont {Xu}, \citenamefont {Yasuda}, \citenamefont {Yoneda}, \citenamefont {Zenihiro}, \citenamefont {Zhou}, \citenamefont {Zuo},\ and\ \citenamefont {Uesaka}}]{17B21}%
  \BibitemOpen
  \bibfield  {author} {\bibinfo {author} {\bibfnamefont {Z.~H.}\ \bibnamefont {Yang}}, \bibinfo {author} {\bibfnamefont {Y.}~\bibnamefont {Kubota}}, \bibinfo {author} {\bibfnamefont {A.}~\bibnamefont {Corsi}}, \bibinfo {author} {\bibfnamefont {K.}~\bibnamefont {Yoshida}}, \bibinfo {author} {\bibfnamefont {X.-X.}\ \bibnamefont {Sun}}, \bibinfo {author} {\bibfnamefont {J.~G.}\ \bibnamefont {Li}}, \bibinfo {author} {\bibfnamefont {M.}~\bibnamefont {Kimura}}, \bibinfo {author} {\bibfnamefont {N.}~\bibnamefont {Michel}}, \bibinfo {author} {\bibfnamefont {K.}~\bibnamefont {Ogata}}, \bibinfo {author} {\bibfnamefont {C.~X.}\ \bibnamefont {Yuan}}, \bibinfo {author} {\bibfnamefont {Q.}~\bibnamefont {Yuan}}, \bibinfo {author} {\bibfnamefont {G.}~\bibnamefont {Authelet}}, \bibinfo {author} {\bibfnamefont {H.}~\bibnamefont {Baba}}, \bibinfo {author} {\bibfnamefont {C.}~\bibnamefont {Caesar}}, \bibinfo {author} {\bibfnamefont {D.}~\bibnamefont {Calvet}}, \bibinfo {author} {\bibfnamefont {A.}~\bibnamefont {Delbart}},
  \bibinfo {author} {\bibfnamefont {M.}~\bibnamefont {Dozono}}, \bibinfo {author} {\bibfnamefont {J.}~\bibnamefont {Feng}}, \bibinfo {author} {\bibfnamefont {F.}~\bibnamefont {Flavigny}}, \bibinfo {author} {\bibfnamefont {J.-M.}\ \bibnamefont {Gheller}}, \bibinfo {author} {\bibfnamefont {J.}~\bibnamefont {Gibelin}}, \bibinfo {author} {\bibfnamefont {A.}~\bibnamefont {Giganon}}, \bibinfo {author} {\bibfnamefont {A.}~\bibnamefont {Gillibert}}, \bibinfo {author} {\bibfnamefont {K.}~\bibnamefont {Hasegawa}}, \bibinfo {author} {\bibfnamefont {T.}~\bibnamefont {Isobe}}, \bibinfo {author} {\bibfnamefont {Y.}~\bibnamefont {Kanaya}}, \bibinfo {author} {\bibfnamefont {S.}~\bibnamefont {Kawakami}}, \bibinfo {author} {\bibfnamefont {D.}~\bibnamefont {Kim}}, \bibinfo {author} {\bibfnamefont {Y.}~\bibnamefont {Kiyokawa}}, \bibinfo {author} {\bibfnamefont {M.}~\bibnamefont {Kobayashi}}, \bibinfo {author} {\bibfnamefont {N.}~\bibnamefont {Kobayashi}}, \bibinfo {author} {\bibfnamefont {T.}~\bibnamefont {Kobayashi}}, \bibinfo
  {author} {\bibfnamefont {Y.}~\bibnamefont {Kondo}}, \bibinfo {author} {\bibfnamefont {Z.}~\bibnamefont {Korkulu}}, \bibinfo {author} {\bibfnamefont {S.}~\bibnamefont {Koyama}}, \bibinfo {author} {\bibfnamefont {V.}~\bibnamefont {Lapoux}}, \bibinfo {author} {\bibfnamefont {Y.}~\bibnamefont {Maeda}}, \bibinfo {author} {\bibfnamefont {F.~M.}\ \bibnamefont {Marqu\'es}}, \bibinfo {author} {\bibfnamefont {T.}~\bibnamefont {Motobayashi}}, \bibinfo {author} {\bibfnamefont {T.}~\bibnamefont {Miyazaki}}, \bibinfo {author} {\bibfnamefont {T.}~\bibnamefont {Nakamura}}, \bibinfo {author} {\bibfnamefont {N.}~\bibnamefont {Nakatsuka}}, \bibinfo {author} {\bibfnamefont {Y.}~\bibnamefont {Nishio}}, \bibinfo {author} {\bibfnamefont {A.}~\bibnamefont {Obertelli}}, \bibinfo {author} {\bibfnamefont {A.}~\bibnamefont {Ohkura}}, \bibinfo {author} {\bibfnamefont {N.~A.}\ \bibnamefont {Orr}}, \bibinfo {author} {\bibfnamefont {S.}~\bibnamefont {Ota}}, \bibinfo {author} {\bibfnamefont {H.}~\bibnamefont {Otsu}}, \bibinfo {author}
  {\bibfnamefont {T.}~\bibnamefont {Ozaki}}, \bibinfo {author} {\bibfnamefont {V.}~\bibnamefont {Panin}}, \bibinfo {author} {\bibfnamefont {S.}~\bibnamefont {Paschalis}}, \bibinfo {author} {\bibfnamefont {E.~C.}\ \bibnamefont {Pollacco}}, \bibinfo {author} {\bibfnamefont {S.}~\bibnamefont {Reichert}}, \bibinfo {author} {\bibfnamefont {J.-Y.}\ \bibnamefont {Rouss\'e}}, \bibinfo {author} {\bibfnamefont {A.~T.}\ \bibnamefont {Saito}}, \bibinfo {author} {\bibfnamefont {S.}~\bibnamefont {Sakaguchi}}, \bibinfo {author} {\bibfnamefont {M.}~\bibnamefont {Sako}}, \bibinfo {author} {\bibfnamefont {C.}~\bibnamefont {Santamaria}}, \bibinfo {author} {\bibfnamefont {M.}~\bibnamefont {Sasano}}, \bibinfo {author} {\bibfnamefont {H.}~\bibnamefont {Sato}}, \bibinfo {author} {\bibfnamefont {M.}~\bibnamefont {Shikata}}, \bibinfo {author} {\bibfnamefont {Y.}~\bibnamefont {Shimizu}}, \bibinfo {author} {\bibfnamefont {Y.}~\bibnamefont {Shindo}}, \bibinfo {author} {\bibfnamefont {L.}~\bibnamefont {Stuhl}}, \bibinfo {author}
  {\bibfnamefont {T.}~\bibnamefont {Sumikama}}, \bibinfo {author} {\bibfnamefont {Y.~L.}\ \bibnamefont {Sun}}, \bibinfo {author} {\bibfnamefont {M.}~\bibnamefont {Tabata}}, \bibinfo {author} {\bibfnamefont {Y.}~\bibnamefont {Togano}}, \bibinfo {author} {\bibfnamefont {J.}~\bibnamefont {Tsubota}}, \bibinfo {author} {\bibfnamefont {F.~R.}\ \bibnamefont {Xu}}, \bibinfo {author} {\bibfnamefont {J.}~\bibnamefont {Yasuda}}, \bibinfo {author} {\bibfnamefont {K.}~\bibnamefont {Yoneda}}, \bibinfo {author} {\bibfnamefont {J.}~\bibnamefont {Zenihiro}}, \bibinfo {author} {\bibfnamefont {S.-G.}\ \bibnamefont {Zhou}}, \bibinfo {author} {\bibfnamefont {W.}~\bibnamefont {Zuo}},\ and\ \bibinfo {author} {\bibfnamefont {T.}~\bibnamefont {Uesaka}},\ }\href {https://doi.org/10.1103/PhysRevLett.126.082501} {\bibfield  {journal} {\bibinfo  {journal} {Phys. Rev. Lett.}\ }\textbf {\bibinfo {volume} {126}},\ \bibinfo {pages} {082501} (\bibinfo {year} {2021})}\BibitemShut {NoStop}%
\bibitem [{\citenamefont {Cook}\ \emph {et~al.}(2020)\citenamefont {Cook}, \citenamefont {Nakamura}, \citenamefont {Kondo}, \citenamefont {Hagino}, \citenamefont {Ogata}, \citenamefont {Saito}, \citenamefont {Achouri}, \citenamefont {Aumann}, \citenamefont {Baba}, \citenamefont {Delaunay}, \citenamefont {Deshayes}, \citenamefont {Doornenbal}, \citenamefont {Fukuda}, \citenamefont {Gibelin}, \citenamefont {Hwang}, \citenamefont {Inabe}, \citenamefont {Isobe}, \citenamefont {Kameda}, \citenamefont {Kanno}, \citenamefont {Kim}, \citenamefont {Kobayashi}, \citenamefont {Kobayashi}, \citenamefont {Kubo}, \citenamefont {Leblond}, \citenamefont {Lee}, \citenamefont {Marqu\'es}, \citenamefont {Minakata}, \citenamefont {Motobayashi}, \citenamefont {Muto}, \citenamefont {Murakami}, \citenamefont {Murai}, \citenamefont {Nakashima}, \citenamefont {Nakatsuka}, \citenamefont {Navin}, \citenamefont {Nishi}, \citenamefont {Ogoshi}, \citenamefont {Orr}, \citenamefont {Otsu}, \citenamefont {Sato}, \citenamefont {Satou},
  \citenamefont {Shimizu}, \citenamefont {Suzuki}, \citenamefont {Takahashi}, \citenamefont {Takeda}, \citenamefont {Takeuchi}, \citenamefont {Tanaka}, \citenamefont {Togano}, \citenamefont {Tsubota}, \citenamefont {Tuff}, \citenamefont {Vandebrouck},\ and\ \citenamefont {Yoneda}}]{19B20}%
  \BibitemOpen
  \bibfield  {author} {\bibinfo {author} {\bibfnamefont {K.~J.}\ \bibnamefont {Cook}}, \bibinfo {author} {\bibfnamefont {T.}~\bibnamefont {Nakamura}}, \bibinfo {author} {\bibfnamefont {Y.}~\bibnamefont {Kondo}}, \bibinfo {author} {\bibfnamefont {K.}~\bibnamefont {Hagino}}, \bibinfo {author} {\bibfnamefont {K.}~\bibnamefont {Ogata}}, \bibinfo {author} {\bibfnamefont {A.~T.}\ \bibnamefont {Saito}}, \bibinfo {author} {\bibfnamefont {N.~L.}\ \bibnamefont {Achouri}}, \bibinfo {author} {\bibfnamefont {T.}~\bibnamefont {Aumann}}, \bibinfo {author} {\bibfnamefont {H.}~\bibnamefont {Baba}}, \bibinfo {author} {\bibfnamefont {F.}~\bibnamefont {Delaunay}}, \bibinfo {author} {\bibfnamefont {Q.}~\bibnamefont {Deshayes}}, \bibinfo {author} {\bibfnamefont {P.}~\bibnamefont {Doornenbal}}, \bibinfo {author} {\bibfnamefont {N.}~\bibnamefont {Fukuda}}, \bibinfo {author} {\bibfnamefont {J.}~\bibnamefont {Gibelin}}, \bibinfo {author} {\bibfnamefont {J.~W.}\ \bibnamefont {Hwang}}, \bibinfo {author} {\bibfnamefont {N.}~\bibnamefont
  {Inabe}}, \bibinfo {author} {\bibfnamefont {T.}~\bibnamefont {Isobe}}, \bibinfo {author} {\bibfnamefont {D.}~\bibnamefont {Kameda}}, \bibinfo {author} {\bibfnamefont {D.}~\bibnamefont {Kanno}}, \bibinfo {author} {\bibfnamefont {S.}~\bibnamefont {Kim}}, \bibinfo {author} {\bibfnamefont {N.}~\bibnamefont {Kobayashi}}, \bibinfo {author} {\bibfnamefont {T.}~\bibnamefont {Kobayashi}}, \bibinfo {author} {\bibfnamefont {T.}~\bibnamefont {Kubo}}, \bibinfo {author} {\bibfnamefont {S.}~\bibnamefont {Leblond}}, \bibinfo {author} {\bibfnamefont {J.}~\bibnamefont {Lee}}, \bibinfo {author} {\bibfnamefont {F.~M.}\ \bibnamefont {Marqu\'es}}, \bibinfo {author} {\bibfnamefont {R.}~\bibnamefont {Minakata}}, \bibinfo {author} {\bibfnamefont {T.}~\bibnamefont {Motobayashi}}, \bibinfo {author} {\bibfnamefont {K.}~\bibnamefont {Muto}}, \bibinfo {author} {\bibfnamefont {T.}~\bibnamefont {Murakami}}, \bibinfo {author} {\bibfnamefont {D.}~\bibnamefont {Murai}}, \bibinfo {author} {\bibfnamefont {T.}~\bibnamefont {Nakashima}},
  \bibinfo {author} {\bibfnamefont {N.}~\bibnamefont {Nakatsuka}}, \bibinfo {author} {\bibfnamefont {A.}~\bibnamefont {Navin}}, \bibinfo {author} {\bibfnamefont {S.}~\bibnamefont {Nishi}}, \bibinfo {author} {\bibfnamefont {S.}~\bibnamefont {Ogoshi}}, \bibinfo {author} {\bibfnamefont {N.~A.}\ \bibnamefont {Orr}}, \bibinfo {author} {\bibfnamefont {H.}~\bibnamefont {Otsu}}, \bibinfo {author} {\bibfnamefont {H.}~\bibnamefont {Sato}}, \bibinfo {author} {\bibfnamefont {Y.}~\bibnamefont {Satou}}, \bibinfo {author} {\bibfnamefont {Y.}~\bibnamefont {Shimizu}}, \bibinfo {author} {\bibfnamefont {H.}~\bibnamefont {Suzuki}}, \bibinfo {author} {\bibfnamefont {K.}~\bibnamefont {Takahashi}}, \bibinfo {author} {\bibfnamefont {H.}~\bibnamefont {Takeda}}, \bibinfo {author} {\bibfnamefont {S.}~\bibnamefont {Takeuchi}}, \bibinfo {author} {\bibfnamefont {R.}~\bibnamefont {Tanaka}}, \bibinfo {author} {\bibfnamefont {Y.}~\bibnamefont {Togano}}, \bibinfo {author} {\bibfnamefont {J.}~\bibnamefont {Tsubota}}, \bibinfo {author}
  {\bibfnamefont {A.~G.}\ \bibnamefont {Tuff}}, \bibinfo {author} {\bibfnamefont {M.}~\bibnamefont {Vandebrouck}},\ and\ \bibinfo {author} {\bibfnamefont {K.}~\bibnamefont {Yoneda}},\ }\href {https://doi.org/10.1103/PhysRevLett.124.212503} {\bibfield  {journal} {\bibinfo  {journal} {Phys. Rev. Lett.}\ }\textbf {\bibinfo {volume} {124}},\ \bibinfo {pages} {212503} (\bibinfo {year} {2020})}\BibitemShut {NoStop}%
\bibitem [{\citenamefont {Tanaka}\ \emph {et~al.}(2010)\citenamefont {Tanaka}, \citenamefont {Yamaguchi}, \citenamefont {Suzuki}, \citenamefont {Ohtsubo}, \citenamefont {Fukuda}, \citenamefont {Nishimura}, \citenamefont {Takechi}, \citenamefont {Ogata}, \citenamefont {Ozawa}, \citenamefont {Izumikawa}, \citenamefont {Aiba}, \citenamefont {Aoi}, \citenamefont {Baba}, \citenamefont {Hashizume}, \citenamefont {Inafuku}, \citenamefont {Iwasa}, \citenamefont {Kobayashi}, \citenamefont {Komuro}, \citenamefont {Kondo}, \citenamefont {Kubo}, \citenamefont {Kurokawa}, \citenamefont {Matsuyama}, \citenamefont {Michimasa}, \citenamefont {Motobayashi}, \citenamefont {Nakabayashi}, \citenamefont {Nakajima}, \citenamefont {Nakamura}, \citenamefont {Sakurai}, \citenamefont {Shinoda}, \citenamefont {Shinohara}, \citenamefont {Suzuki}, \citenamefont {Takeshita}, \citenamefont {Takeuchi}, \citenamefont {Togano}, \citenamefont {Yamada}, \citenamefont {Yasuno},\ and\ \citenamefont {Yoshitake}}]{Tanaka2010}%
  \BibitemOpen
  \bibfield  {author} {\bibinfo {author} {\bibfnamefont {K.}~\bibnamefont {Tanaka}}, \bibinfo {author} {\bibfnamefont {T.}~\bibnamefont {Yamaguchi}}, \bibinfo {author} {\bibfnamefont {T.}~\bibnamefont {Suzuki}}, \bibinfo {author} {\bibfnamefont {T.}~\bibnamefont {Ohtsubo}}, \bibinfo {author} {\bibfnamefont {M.}~\bibnamefont {Fukuda}}, \bibinfo {author} {\bibfnamefont {D.}~\bibnamefont {Nishimura}}, \bibinfo {author} {\bibfnamefont {M.}~\bibnamefont {Takechi}}, \bibinfo {author} {\bibfnamefont {K.}~\bibnamefont {Ogata}}, \bibinfo {author} {\bibfnamefont {A.}~\bibnamefont {Ozawa}}, \bibinfo {author} {\bibfnamefont {T.}~\bibnamefont {Izumikawa}}, \bibinfo {author} {\bibfnamefont {T.}~\bibnamefont {Aiba}}, \bibinfo {author} {\bibfnamefont {N.}~\bibnamefont {Aoi}}, \bibinfo {author} {\bibfnamefont {H.}~\bibnamefont {Baba}}, \bibinfo {author} {\bibfnamefont {Y.}~\bibnamefont {Hashizume}}, \bibinfo {author} {\bibfnamefont {K.}~\bibnamefont {Inafuku}}, \bibinfo {author} {\bibfnamefont {N.}~\bibnamefont {Iwasa}},
  \bibinfo {author} {\bibfnamefont {K.}~\bibnamefont {Kobayashi}}, \bibinfo {author} {\bibfnamefont {M.}~\bibnamefont {Komuro}}, \bibinfo {author} {\bibfnamefont {Y.}~\bibnamefont {Kondo}}, \bibinfo {author} {\bibfnamefont {T.}~\bibnamefont {Kubo}}, \bibinfo {author} {\bibfnamefont {M.}~\bibnamefont {Kurokawa}}, \bibinfo {author} {\bibfnamefont {T.}~\bibnamefont {Matsuyama}}, \bibinfo {author} {\bibfnamefont {S.}~\bibnamefont {Michimasa}}, \bibinfo {author} {\bibfnamefont {T.}~\bibnamefont {Motobayashi}}, \bibinfo {author} {\bibfnamefont {T.}~\bibnamefont {Nakabayashi}}, \bibinfo {author} {\bibfnamefont {S.}~\bibnamefont {Nakajima}}, \bibinfo {author} {\bibfnamefont {T.}~\bibnamefont {Nakamura}}, \bibinfo {author} {\bibfnamefont {H.}~\bibnamefont {Sakurai}}, \bibinfo {author} {\bibfnamefont {R.}~\bibnamefont {Shinoda}}, \bibinfo {author} {\bibfnamefont {M.}~\bibnamefont {Shinohara}}, \bibinfo {author} {\bibfnamefont {H.}~\bibnamefont {Suzuki}}, \bibinfo {author} {\bibfnamefont {E.}~\bibnamefont {Takeshita}},
  \bibinfo {author} {\bibfnamefont {S.}~\bibnamefont {Takeuchi}}, \bibinfo {author} {\bibfnamefont {Y.}~\bibnamefont {Togano}}, \bibinfo {author} {\bibfnamefont {K.}~\bibnamefont {Yamada}}, \bibinfo {author} {\bibfnamefont {T.}~\bibnamefont {Yasuno}},\ and\ \bibinfo {author} {\bibfnamefont {M.}~\bibnamefont {Yoshitake}},\ }\href {https://doi.org/10.1103/PhysRevLett.104.062701} {\bibfield  {journal} {\bibinfo  {journal} {Phys. Rev. Lett.}\ }\textbf {\bibinfo {volume} {104}},\ \bibinfo {pages} {062701} (\bibinfo {year} {2010})}\BibitemShut {NoStop}%
\bibitem [{\citenamefont {Bagchi}\ \emph {et~al.}(2020)\citenamefont {Bagchi}, \citenamefont {Kanungo}, \citenamefont {Tanaka}, \citenamefont {Geissel}, \citenamefont {Doornenbal}, \citenamefont {Horiuchi}, \citenamefont {Hagen}, \citenamefont {Suzuki}, \citenamefont {Tsunoda}, \citenamefont {Ahn}, \citenamefont {Baba}, \citenamefont {Behr}, \citenamefont {Browne}, \citenamefont {Chen}, \citenamefont {Cort\'es}, \citenamefont {Estrad\'e}, \citenamefont {Fukuda}, \citenamefont {Holl}, \citenamefont {Itahashi}, \citenamefont {Iwasa}, \citenamefont {Jansen}, \citenamefont {Jiang}, \citenamefont {Kaur}, \citenamefont {Macchiavelli}, \citenamefont {Matsumoto}, \citenamefont {Momiyama}, \citenamefont {Murray}, \citenamefont {Nakamura}, \citenamefont {Novario}, \citenamefont {Ong}, \citenamefont {Otsuka}, \citenamefont {Papenbrock}, \citenamefont {Paschalis}, \citenamefont {Prochazka}, \citenamefont {Scheidenberger}, \citenamefont {Schrock}, \citenamefont {Shimizu}, \citenamefont {Steppenbeck}, \citenamefont
  {Sakurai}, \citenamefont {Suzuki}, \citenamefont {Suzuki}, \citenamefont {Takechi}, \citenamefont {Takeda}, \citenamefont {Takeuchi}, \citenamefont {Taniuchi}, \citenamefont {Wimmer},\ and\ \citenamefont {Yoshida}}]{29F20}%
  \BibitemOpen
  \bibfield  {author} {\bibinfo {author} {\bibfnamefont {S.}~\bibnamefont {Bagchi}}, \bibinfo {author} {\bibfnamefont {R.}~\bibnamefont {Kanungo}}, \bibinfo {author} {\bibfnamefont {Y.~K.}\ \bibnamefont {Tanaka}}, \bibinfo {author} {\bibfnamefont {H.}~\bibnamefont {Geissel}}, \bibinfo {author} {\bibfnamefont {P.}~\bibnamefont {Doornenbal}}, \bibinfo {author} {\bibfnamefont {W.}~\bibnamefont {Horiuchi}}, \bibinfo {author} {\bibfnamefont {G.}~\bibnamefont {Hagen}}, \bibinfo {author} {\bibfnamefont {T.}~\bibnamefont {Suzuki}}, \bibinfo {author} {\bibfnamefont {N.}~\bibnamefont {Tsunoda}}, \bibinfo {author} {\bibfnamefont {D.~S.}\ \bibnamefont {Ahn}}, \bibinfo {author} {\bibfnamefont {H.}~\bibnamefont {Baba}}, \bibinfo {author} {\bibfnamefont {K.}~\bibnamefont {Behr}}, \bibinfo {author} {\bibfnamefont {F.}~\bibnamefont {Browne}}, \bibinfo {author} {\bibfnamefont {S.}~\bibnamefont {Chen}}, \bibinfo {author} {\bibfnamefont {M.~L.}\ \bibnamefont {Cort\'es}}, \bibinfo {author} {\bibfnamefont {A.}~\bibnamefont
  {Estrad\'e}}, \bibinfo {author} {\bibfnamefont {N.}~\bibnamefont {Fukuda}}, \bibinfo {author} {\bibfnamefont {M.}~\bibnamefont {Holl}}, \bibinfo {author} {\bibfnamefont {K.}~\bibnamefont {Itahashi}}, \bibinfo {author} {\bibfnamefont {N.}~\bibnamefont {Iwasa}}, \bibinfo {author} {\bibfnamefont {G.~R.}\ \bibnamefont {Jansen}}, \bibinfo {author} {\bibfnamefont {W.~G.}\ \bibnamefont {Jiang}}, \bibinfo {author} {\bibfnamefont {S.}~\bibnamefont {Kaur}}, \bibinfo {author} {\bibfnamefont {A.~O.}\ \bibnamefont {Macchiavelli}}, \bibinfo {author} {\bibfnamefont {S.~Y.}\ \bibnamefont {Matsumoto}}, \bibinfo {author} {\bibfnamefont {S.}~\bibnamefont {Momiyama}}, \bibinfo {author} {\bibfnamefont {I.}~\bibnamefont {Murray}}, \bibinfo {author} {\bibfnamefont {T.}~\bibnamefont {Nakamura}}, \bibinfo {author} {\bibfnamefont {S.~J.}\ \bibnamefont {Novario}}, \bibinfo {author} {\bibfnamefont {H.~J.}\ \bibnamefont {Ong}}, \bibinfo {author} {\bibfnamefont {T.}~\bibnamefont {Otsuka}}, \bibinfo {author} {\bibfnamefont
  {T.}~\bibnamefont {Papenbrock}}, \bibinfo {author} {\bibfnamefont {S.}~\bibnamefont {Paschalis}}, \bibinfo {author} {\bibfnamefont {A.}~\bibnamefont {Prochazka}}, \bibinfo {author} {\bibfnamefont {C.}~\bibnamefont {Scheidenberger}}, \bibinfo {author} {\bibfnamefont {P.}~\bibnamefont {Schrock}}, \bibinfo {author} {\bibfnamefont {Y.}~\bibnamefont {Shimizu}}, \bibinfo {author} {\bibfnamefont {D.}~\bibnamefont {Steppenbeck}}, \bibinfo {author} {\bibfnamefont {H.}~\bibnamefont {Sakurai}}, \bibinfo {author} {\bibfnamefont {D.}~\bibnamefont {Suzuki}}, \bibinfo {author} {\bibfnamefont {H.}~\bibnamefont {Suzuki}}, \bibinfo {author} {\bibfnamefont {M.}~\bibnamefont {Takechi}}, \bibinfo {author} {\bibfnamefont {H.}~\bibnamefont {Takeda}}, \bibinfo {author} {\bibfnamefont {S.}~\bibnamefont {Takeuchi}}, \bibinfo {author} {\bibfnamefont {R.}~\bibnamefont {Taniuchi}}, \bibinfo {author} {\bibfnamefont {K.}~\bibnamefont {Wimmer}},\ and\ \bibinfo {author} {\bibfnamefont {K.}~\bibnamefont {Yoshida}},\ }\href
  {https://doi.org/10.1103/PhysRevLett.124.222504} {\bibfield  {journal} {\bibinfo  {journal} {Phys. Rev. Lett.}\ }\textbf {\bibinfo {volume} {124}},\ \bibinfo {pages} {222504} (\bibinfo {year} {2020})}\BibitemShut {NoStop}%
\bibitem [{\citenamefont {Sagawa}(1992)}]{SAGAWA1992}%
  \BibitemOpen
  \bibfield  {author} {\bibinfo {author} {\bibfnamefont {H.}~\bibnamefont {Sagawa}},\ }\href {https://doi.org/https://doi.org/10.1016/0370-2693(92)90150-3} {\bibfield  {journal} {\bibinfo  {journal} {Phys. Lett. B}\ }\textbf {\bibinfo {volume} {286}},\ \bibinfo {pages} {7} (\bibinfo {year} {1992})}\BibitemShut {NoStop}%
\bibitem [{\citenamefont {Riisager}\ \emph {et~al.}(1992)\citenamefont {Riisager}, \citenamefont {Jensen},\ and\ \citenamefont {M{\o}ller}}]{RIISAGER1992}%
  \BibitemOpen
  \bibfield  {author} {\bibinfo {author} {\bibfnamefont {K.}~\bibnamefont {Riisager}}, \bibinfo {author} {\bibfnamefont {A.}~\bibnamefont {Jensen}},\ and\ \bibinfo {author} {\bibfnamefont {P.}~\bibnamefont {M{\o}ller}},\ }\href {https://doi.org/https://doi.org/10.1016/0375-9474(92)90691-C} {\bibfield  {journal} {\bibinfo  {journal} {Nucl. Phys. A}\ }\textbf {\bibinfo {volume} {548}},\ \bibinfo {pages} {393} (\bibinfo {year} {1992})}\BibitemShut {NoStop}%
\bibitem [{\citenamefont {Meng}\ and\ \citenamefont {Ring}(1996)}]{meng96}%
  \BibitemOpen
  \bibfield  {author} {\bibinfo {author} {\bibfnamefont {J.}~\bibnamefont {Meng}}\ and\ \bibinfo {author} {\bibfnamefont {P.}~\bibnamefont {Ring}},\ }\href {https://doi.org/10.1103/PhysRevLett.77.3963} {\bibfield  {journal} {\bibinfo  {journal} {Phys. Rev. Lett.}\ }\textbf {\bibinfo {volume} {77}},\ \bibinfo {pages} {3963} (\bibinfo {year} {1996})}\BibitemShut {NoStop}%
\bibitem [{\citenamefont {Meng}\ \emph {et~al.}(2006)\citenamefont {Meng}, \citenamefont {Toki}, \citenamefont {Zhou}, \citenamefont {Zhang}, \citenamefont {Long},\ and\ \citenamefont {Geng}}]{MENG06}%
  \BibitemOpen
  \bibfield  {author} {\bibinfo {author} {\bibfnamefont {J.}~\bibnamefont {Meng}}, \bibinfo {author} {\bibfnamefont {H.}~\bibnamefont {Toki}}, \bibinfo {author} {\bibfnamefont {S.}~\bibnamefont {Zhou}}, \bibinfo {author} {\bibfnamefont {S.}~\bibnamefont {Zhang}}, \bibinfo {author} {\bibfnamefont {W.}~\bibnamefont {Long}},\ and\ \bibinfo {author} {\bibfnamefont {L.}~\bibnamefont {Geng}},\ }\href {https://doi.org/https://doi.org/10.1016/j.ppnp.2005.06.001} {\bibfield  {journal} {\bibinfo  {journal} {Prog. Part. Nucl. Phys.}\ }\textbf {\bibinfo {volume} {57}},\ \bibinfo {pages} {470} (\bibinfo {year} {2006})}\BibitemShut {NoStop}%
\bibitem [{\citenamefont {Meng}\ and\ \citenamefont {Zhou}(2015)}]{Meng_2015}%
  \BibitemOpen
  \bibfield  {author} {\bibinfo {author} {\bibfnamefont {J.}~\bibnamefont {Meng}}\ and\ \bibinfo {author} {\bibfnamefont {S.~G.}\ \bibnamefont {Zhou}},\ }\href {https://doi.org/10.1088/0954-3899/42/9/093101} {\bibfield  {journal} {\bibinfo  {journal} {J. Phys. G: Nucl. Part. Phys.}\ }\textbf {\bibinfo {volume} {42}},\ \bibinfo {pages} {093101} (\bibinfo {year} {2015})}\BibitemShut {NoStop}%
\bibitem [{\citenamefont {Bohr}\ and\ \citenamefont {Mottelson}(1975)}]{BM2}%
  \BibitemOpen
  \bibfield  {author} {\bibinfo {author} {\bibfnamefont {A.}~\bibnamefont {Bohr}}\ and\ \bibinfo {author} {\bibfnamefont {B.~R.}\ \bibnamefont {Mottelson}},\ }\href@noop {} {\bibfield  {journal} {\bibinfo  {journal} {Nuclear Struture Vol. II}\ } (\bibinfo {year} {1975})}\BibitemShut {NoStop}%
\bibitem [{\citenamefont {Hamamoto}(2005)}]{Hamamoto2005}%
  \BibitemOpen
  \bibfield  {author} {\bibinfo {author} {\bibfnamefont {I.}~\bibnamefont {Hamamoto}},\ }\href {https://doi.org/10.1103/PhysRevC.72.024301} {\bibfield  {journal} {\bibinfo  {journal} {Phys. Rev. C}\ }\textbf {\bibinfo {volume} {72}},\ \bibinfo {pages} {024301} (\bibinfo {year} {2005})}\BibitemShut {NoStop}%
\bibitem [{\citenamefont {Nakamura}\ \emph {et~al.}(1997)\citenamefont {Nakamura}, \citenamefont {Motobayashi}, \citenamefont {Ando}, \citenamefont {Mengoni}, \citenamefont {Nishio}, \citenamefont {Sakurai}, \citenamefont {Shimoura}, \citenamefont {Teranishi}, \citenamefont {Yanagisawa},\ and\ \citenamefont {Ishihara}}]{NAKAMURA1997}%
  \BibitemOpen
  \bibfield  {author} {\bibinfo {author} {\bibfnamefont {T.}~\bibnamefont {Nakamura}}, \bibinfo {author} {\bibfnamefont {T.}~\bibnamefont {Motobayashi}}, \bibinfo {author} {\bibfnamefont {Y.}~\bibnamefont {Ando}}, \bibinfo {author} {\bibfnamefont {A.}~\bibnamefont {Mengoni}}, \bibinfo {author} {\bibfnamefont {T.}~\bibnamefont {Nishio}}, \bibinfo {author} {\bibfnamefont {H.}~\bibnamefont {Sakurai}}, \bibinfo {author} {\bibfnamefont {S.}~\bibnamefont {Shimoura}}, \bibinfo {author} {\bibfnamefont {T.}~\bibnamefont {Teranishi}}, \bibinfo {author} {\bibfnamefont {Y.}~\bibnamefont {Yanagisawa}},\ and\ \bibinfo {author} {\bibfnamefont {M.}~\bibnamefont {Ishihara}},\ }\href {https://doi.org/https://doi.org/10.1016/S0370-2693(96)01690-5} {\bibfield  {journal} {\bibinfo  {journal} {Phys. Lett. B}\ }\textbf {\bibinfo {volume} {394}},\ \bibinfo {pages} {11} (\bibinfo {year} {1997})}\BibitemShut {NoStop}%
\bibitem [{\citenamefont {Nakamura}\ \emph {et~al.}(2006)\citenamefont {Nakamura}, \citenamefont {Vinodkumar}, \citenamefont {Sugimoto}, \citenamefont {Aoi}, \citenamefont {Baba}, \citenamefont {Bazin}, \citenamefont {Fukuda}, \citenamefont {Gomi}, \citenamefont {Hasegawa}, \citenamefont {Imai}, \citenamefont {Ishihara}, \citenamefont {Kobayashi}, \citenamefont {Kondo}, \citenamefont {Kubo}, \citenamefont {Miura}, \citenamefont {Motobayashi}, \citenamefont {Otsu}, \citenamefont {Saito}, \citenamefont {Sakurai}, \citenamefont {Shimoura}, \citenamefont {Watanabe}, \citenamefont {Watanabe}, \citenamefont {Yakushiji}, \citenamefont {Yanagisawa},\ and\ \citenamefont {Yoneda}}]{Nakamura2006}%
  \BibitemOpen
  \bibfield  {author} {\bibinfo {author} {\bibfnamefont {T.}~\bibnamefont {Nakamura}}, \bibinfo {author} {\bibfnamefont {A.~M.}\ \bibnamefont {Vinodkumar}}, \bibinfo {author} {\bibfnamefont {T.}~\bibnamefont {Sugimoto}}, \bibinfo {author} {\bibfnamefont {N.}~\bibnamefont {Aoi}}, \bibinfo {author} {\bibfnamefont {H.}~\bibnamefont {Baba}}, \bibinfo {author} {\bibfnamefont {D.}~\bibnamefont {Bazin}}, \bibinfo {author} {\bibfnamefont {N.}~\bibnamefont {Fukuda}}, \bibinfo {author} {\bibfnamefont {T.}~\bibnamefont {Gomi}}, \bibinfo {author} {\bibfnamefont {H.}~\bibnamefont {Hasegawa}}, \bibinfo {author} {\bibfnamefont {N.}~\bibnamefont {Imai}}, \bibinfo {author} {\bibfnamefont {M.}~\bibnamefont {Ishihara}}, \bibinfo {author} {\bibfnamefont {T.}~\bibnamefont {Kobayashi}}, \bibinfo {author} {\bibfnamefont {Y.}~\bibnamefont {Kondo}}, \bibinfo {author} {\bibfnamefont {T.}~\bibnamefont {Kubo}}, \bibinfo {author} {\bibfnamefont {M.}~\bibnamefont {Miura}}, \bibinfo {author} {\bibfnamefont {T.}~\bibnamefont {Motobayashi}},
  \bibinfo {author} {\bibfnamefont {H.}~\bibnamefont {Otsu}}, \bibinfo {author} {\bibfnamefont {A.}~\bibnamefont {Saito}}, \bibinfo {author} {\bibfnamefont {H.}~\bibnamefont {Sakurai}}, \bibinfo {author} {\bibfnamefont {S.}~\bibnamefont {Shimoura}}, \bibinfo {author} {\bibfnamefont {K.}~\bibnamefont {Watanabe}}, \bibinfo {author} {\bibfnamefont {Y.~X.}\ \bibnamefont {Watanabe}}, \bibinfo {author} {\bibfnamefont {T.}~\bibnamefont {Yakushiji}}, \bibinfo {author} {\bibfnamefont {Y.}~\bibnamefont {Yanagisawa}},\ and\ \bibinfo {author} {\bibfnamefont {K.}~\bibnamefont {Yoneda}},\ }\href {https://doi.org/10.1103/PhysRevLett.96.252502} {\bibfield  {journal} {\bibinfo  {journal} {Phys. Rev. Lett.}\ }\textbf {\bibinfo {volume} {96}},\ \bibinfo {pages} {252502} (\bibinfo {year} {2006})}\BibitemShut {NoStop}%
\bibitem [{\citenamefont {Misu}\ \emph {et~al.}(1997)\citenamefont {Misu}, \citenamefont {Nazarewicz},\ and\ \citenamefont {{\AA}berg}}]{MISU1997}%
  \BibitemOpen
  \bibfield  {author} {\bibinfo {author} {\bibfnamefont {T.}~\bibnamefont {Misu}}, \bibinfo {author} {\bibfnamefont {W.}~\bibnamefont {Nazarewicz}},\ and\ \bibinfo {author} {\bibfnamefont {S.}~\bibnamefont {{\AA}berg}},\ }\href {https://doi.org/https://doi.org/10.1016/S0375-9474(96)00458-7} {\bibfield  {journal} {\bibinfo  {journal} {Nucl. Phys. A}\ }\textbf {\bibinfo {volume} {614}},\ \bibinfo {pages} {44} (\bibinfo {year} {1997})}\BibitemShut {NoStop}%
\bibitem [{\citenamefont {Hamamoto}(2012)}]{Hamamoto2012}%
  \BibitemOpen
  \bibfield  {author} {\bibinfo {author} {\bibfnamefont {I.}~\bibnamefont {Hamamoto}},\ }\href {https://doi.org/10.1103/PhysRevC.85.064329} {\bibfield  {journal} {\bibinfo  {journal} {Phys. Rev. C}\ }\textbf {\bibinfo {volume} {85}},\ \bibinfo {pages} {064329} (\bibinfo {year} {2012})}\BibitemShut {NoStop}%
\bibitem [{\citenamefont {Hamamoto}(2010)}]{Hamamoto2010}%
  \BibitemOpen
  \bibfield  {author} {\bibinfo {author} {\bibfnamefont {I.}~\bibnamefont {Hamamoto}},\ }\href {https://doi.org/10.1103/PhysRevC.81.021304} {\bibfield  {journal} {\bibinfo  {journal} {Phys. Rev. C}\ }\textbf {\bibinfo {volume} {81}},\ \bibinfo {pages} {021304} (\bibinfo {year} {2010})}\BibitemShut {NoStop}%
\bibitem [{\citenamefont {Urata}\ \emph {et~al.}(2011)\citenamefont {Urata}, \citenamefont {Hagino},\ and\ \citenamefont {Sagawa}}]{Urata2011}%
  \BibitemOpen
  \bibfield  {author} {\bibinfo {author} {\bibfnamefont {Y.}~\bibnamefont {Urata}}, \bibinfo {author} {\bibfnamefont {K.}~\bibnamefont {Hagino}},\ and\ \bibinfo {author} {\bibfnamefont {H.}~\bibnamefont {Sagawa}},\ }\href {https://doi.org/10.1103/PhysRevC.83.041303} {\bibfield  {journal} {\bibinfo  {journal} {Phys. Rev. C}\ }\textbf {\bibinfo {volume} {83}},\ \bibinfo {pages} {041303} (\bibinfo {year} {2011})}\BibitemShut {NoStop}%
\bibitem [{\citenamefont {Urata}\ \emph {et~al.}(2012)\citenamefont {Urata}, \citenamefont {Hagino},\ and\ \citenamefont {Sagawa}}]{Urata2012}%
  \BibitemOpen
  \bibfield  {author} {\bibinfo {author} {\bibfnamefont {Y.}~\bibnamefont {Urata}}, \bibinfo {author} {\bibfnamefont {K.}~\bibnamefont {Hagino}},\ and\ \bibinfo {author} {\bibfnamefont {H.}~\bibnamefont {Sagawa}},\ }\href {https://doi.org/10.1103/PhysRevC.86.044613} {\bibfield  {journal} {\bibinfo  {journal} {Phys. Rev. C}\ }\textbf {\bibinfo {volume} {86}},\ \bibinfo {pages} {044613} (\bibinfo {year} {2012})}\BibitemShut {NoStop}%
\bibitem [{\citenamefont {Urata}\ \emph {et~al.}(2017)\citenamefont {Urata}, \citenamefont {Hagino},\ and\ \citenamefont {Sagawa}}]{Urata2013}%
  \BibitemOpen
  \bibfield  {author} {\bibinfo {author} {\bibfnamefont {Y.}~\bibnamefont {Urata}}, \bibinfo {author} {\bibfnamefont {K.}~\bibnamefont {Hagino}},\ and\ \bibinfo {author} {\bibfnamefont {H.}~\bibnamefont {Sagawa}},\ }\href {https://doi.org/10.1103/PhysRevC.96.064311} {\bibfield  {journal} {\bibinfo  {journal} {Phys. Rev. C}\ }\textbf {\bibinfo {volume} {96}},\ \bibinfo {pages} {064311} (\bibinfo {year} {2017})}\BibitemShut {NoStop}%
\bibitem [{\citenamefont {Otsuka}\ \emph {et~al.}(1993)\citenamefont {Otsuka}, \citenamefont {Fukunishi},\ and\ \citenamefont {Sagawa}}]{Otsuka93}%
  \BibitemOpen
  \bibfield  {author} {\bibinfo {author} {\bibfnamefont {T.}~\bibnamefont {Otsuka}}, \bibinfo {author} {\bibfnamefont {N.}~\bibnamefont {Fukunishi}},\ and\ \bibinfo {author} {\bibfnamefont {H.}~\bibnamefont {Sagawa}},\ }\href {https://doi.org/10.1103/PhysRevLett.70.1385} {\bibfield  {journal} {\bibinfo  {journal} {Phys. Rev. Lett.}\ }\textbf {\bibinfo {volume} {70}},\ \bibinfo {pages} {1385} (\bibinfo {year} {1993})}\BibitemShut {NoStop}%
\bibitem [{\citenamefont {Kuo}\ \emph {et~al.}(1997)\citenamefont {Kuo}, \citenamefont {Krmpoti\ifmmode~\acute{c}\else \'{c}\fi{}},\ and\ \citenamefont {Tzeng}}]{Kuo97}%
  \BibitemOpen
  \bibfield  {author} {\bibinfo {author} {\bibfnamefont {T.~T.~S.}\ \bibnamefont {Kuo}}, \bibinfo {author} {\bibfnamefont {F.}~\bibnamefont {Krmpoti\ifmmode~\acute{c}\else \'{c}\fi{}}},\ and\ \bibinfo {author} {\bibfnamefont {Y.}~\bibnamefont {Tzeng}},\ }\href {https://doi.org/10.1103/PhysRevLett.78.2708} {\bibfield  {journal} {\bibinfo  {journal} {Phys. Rev. Lett.}\ }\textbf {\bibinfo {volume} {78}},\ \bibinfo {pages} {2708} (\bibinfo {year} {1997})}\BibitemShut {NoStop}%
\bibitem [{\citenamefont {Descouvemont}(1999)}]{DESCOUVEMONT1999}%
  \BibitemOpen
  \bibfield  {author} {\bibinfo {author} {\bibfnamefont {P.}~\bibnamefont {Descouvemont}},\ }\href {https://doi.org/https://doi.org/10.1016/S0375-9474(99)00305-X} {\bibfield  {journal} {\bibinfo  {journal} {Nucl. Phys. A}\ }\textbf {\bibinfo {volume} {655}},\ \bibinfo {pages} {440} (\bibinfo {year} {1999})}\BibitemShut {NoStop}%
\bibitem [{\citenamefont {Zhou}\ \emph {et~al.}(2010)\citenamefont {Zhou}, \citenamefont {Meng}, \citenamefont {Ring},\ and\ \citenamefont {Zhao}}]{Zhou10}%
  \BibitemOpen
  \bibfield  {author} {\bibinfo {author} {\bibfnamefont {S.-G.}\ \bibnamefont {Zhou}}, \bibinfo {author} {\bibfnamefont {J.}~\bibnamefont {Meng}}, \bibinfo {author} {\bibfnamefont {P.}~\bibnamefont {Ring}},\ and\ \bibinfo {author} {\bibfnamefont {E.-G.}\ \bibnamefont {Zhao}},\ }\href {https://doi.org/10.1103/PhysRevC.82.011301} {\bibfield  {journal} {\bibinfo  {journal} {Phys. Rev. C}\ }\textbf {\bibinfo {volume} {82}},\ \bibinfo {pages} {011301} (\bibinfo {year} {2010})}\BibitemShut {NoStop}%
\bibitem [{\citenamefont {Li}\ \emph {et~al.}(2012)\citenamefont {Li}, \citenamefont {Meng}, \citenamefont {Ring}, \citenamefont {Zhao},\ and\ \citenamefont {Zhou}}]{li12}%
  \BibitemOpen
  \bibfield  {author} {\bibinfo {author} {\bibfnamefont {L.}~\bibnamefont {Li}}, \bibinfo {author} {\bibfnamefont {J.}~\bibnamefont {Meng}}, \bibinfo {author} {\bibfnamefont {P.}~\bibnamefont {Ring}}, \bibinfo {author} {\bibfnamefont {E.-G.}\ \bibnamefont {Zhao}},\ and\ \bibinfo {author} {\bibfnamefont {S.-G.}\ \bibnamefont {Zhou}},\ }\href {https://doi.org/10.1103/PhysRevC.85.024312} {\bibfield  {journal} {\bibinfo  {journal} {Phys. Rev. C}\ }\textbf {\bibinfo {volume} {85}},\ \bibinfo {pages} {024312} (\bibinfo {year} {2012})}\BibitemShut {NoStop}%
\bibitem [{\citenamefont {Zhang}\ \emph {et~al.}(2013)\citenamefont {Zhang}, \citenamefont {Pei},\ and\ \citenamefont {Xu}}]{pei131}%
  \BibitemOpen
  \bibfield  {author} {\bibinfo {author} {\bibfnamefont {Y.~N.}\ \bibnamefont {Zhang}}, \bibinfo {author} {\bibfnamefont {J.~C.}\ \bibnamefont {Pei}},\ and\ \bibinfo {author} {\bibfnamefont {F.~R.}\ \bibnamefont {Xu}},\ }\href {https://doi.org/10.1103/PhysRevC.88.054305} {\bibfield  {journal} {\bibinfo  {journal} {Phys. Rev. C}\ }\textbf {\bibinfo {volume} {88}},\ \bibinfo {pages} {054305} (\bibinfo {year} {2013})}\BibitemShut {NoStop}%
\bibitem [{\citenamefont {Pei}\ \emph {et~al.}(2013)\citenamefont {Pei}, \citenamefont {Zhang},\ and\ \citenamefont {Xu}}]{pei132}%
  \BibitemOpen
  \bibfield  {author} {\bibinfo {author} {\bibfnamefont {J.~C.}\ \bibnamefont {Pei}}, \bibinfo {author} {\bibfnamefont {Y.~N.}\ \bibnamefont {Zhang}},\ and\ \bibinfo {author} {\bibfnamefont {F.~R.}\ \bibnamefont {Xu}},\ }\href {https://doi.org/10.1103/PhysRevC.87.051302} {\bibfield  {journal} {\bibinfo  {journal} {Phys. Rev. C}\ }\textbf {\bibinfo {volume} {87}},\ \bibinfo {pages} {051302} (\bibinfo {year} {2013})}\BibitemShut {NoStop}%
\bibitem [{\citenamefont {Sun}\ \emph {et~al.}(2018)\citenamefont {Sun}, \citenamefont {Zhao},\ and\ \citenamefont {Zhou}}]{SUN18}%
  \BibitemOpen
  \bibfield  {author} {\bibinfo {author} {\bibfnamefont {X.-X.}\ \bibnamefont {Sun}}, \bibinfo {author} {\bibfnamefont {J.}~\bibnamefont {Zhao}},\ and\ \bibinfo {author} {\bibfnamefont {S.-G.}\ \bibnamefont {Zhou}},\ }\href {https://doi.org/https://doi.org/10.1016/j.physletb.2018.08.071} {\bibfield  {journal} {\bibinfo  {journal} {Phys. Lett. B}\ }\textbf {\bibinfo {volume} {785}},\ \bibinfo {pages} {530} (\bibinfo {year} {2018})}\BibitemShut {NoStop}%
\bibitem [{\citenamefont {Nakada}\ and\ \citenamefont {Takayama}(2018)}]{Nakada18}%
  \BibitemOpen
  \bibfield  {author} {\bibinfo {author} {\bibfnamefont {H.}~\bibnamefont {Nakada}}\ and\ \bibinfo {author} {\bibfnamefont {K.}~\bibnamefont {Takayama}},\ }\href {https://doi.org/10.1103/PhysRevC.98.011301} {\bibfield  {journal} {\bibinfo  {journal} {Phys. Rev. C}\ }\textbf {\bibinfo {volume} {98}},\ \bibinfo {pages} {011301} (\bibinfo {year} {2018})}\BibitemShut {NoStop}%
\bibitem [{\citenamefont {Zhang}\ \emph {et~al.}(2019)\citenamefont {Zhang}, \citenamefont {Wang},\ and\ \citenamefont {Zhang}}]{zhang19}%
  \BibitemOpen
  \bibfield  {author} {\bibinfo {author} {\bibfnamefont {K.~Y.}\ \bibnamefont {Zhang}}, \bibinfo {author} {\bibfnamefont {D.~Y.}\ \bibnamefont {Wang}},\ and\ \bibinfo {author} {\bibfnamefont {S.~Q.}\ \bibnamefont {Zhang}},\ }\href {https://doi.org/10.1103/PhysRevC.100.034312} {\bibfield  {journal} {\bibinfo  {journal} {Phys. Rev. C}\ }\textbf {\bibinfo {volume} {100}},\ \bibinfo {pages} {034312} (\bibinfo {year} {2019})}\BibitemShut {NoStop}%
\bibitem [{\citenamefont {Sun}\ \emph {et~al.}(2020)\citenamefont {Sun}, \citenamefont {Zhao},\ and\ \citenamefont {Zhou}}]{SUN20}%
  \BibitemOpen
  \bibfield  {author} {\bibinfo {author} {\bibfnamefont {X.-X.}\ \bibnamefont {Sun}}, \bibinfo {author} {\bibfnamefont {J.}~\bibnamefont {Zhao}},\ and\ \bibinfo {author} {\bibfnamefont {S.-G.}\ \bibnamefont {Zhou}},\ }\href {https://doi.org/https://doi.org/10.1016/j.nuclphysa.2020.122011} {\bibfield  {journal} {\bibinfo  {journal} {Nucl. Phys. A}\ }\textbf {\bibinfo {volume} {1003}},\ \bibinfo {pages} {122011} (\bibinfo {year} {2020})}\BibitemShut {NoStop}%
\bibitem [{\citenamefont {Sun}(2021)}]{SUN21}%
  \BibitemOpen
  \bibfield  {author} {\bibinfo {author} {\bibfnamefont {X.-X.}\ \bibnamefont {Sun}},\ }\href {https://doi.org/10.1103/PhysRevC.103.054315} {\bibfield  {journal} {\bibinfo  {journal} {Phys. Rev. C}\ }\textbf {\bibinfo {volume} {103}},\ \bibinfo {pages} {054315} (\bibinfo {year} {2021})}\BibitemShut {NoStop}%
\bibitem [{\citenamefont {Sun}\ and\ \citenamefont {Zhou}(2021)}]{SUN20212072}%
  \BibitemOpen
  \bibfield  {author} {\bibinfo {author} {\bibfnamefont {X.-X.}\ \bibnamefont {Sun}}\ and\ \bibinfo {author} {\bibfnamefont {S.-G.}\ \bibnamefont {Zhou}},\ }\href {https://doi.org/https://doi.org/10.1016/j.scib.2021.07.005} {\bibfield  {journal} {\bibinfo  {journal} {Sci. Bull.}\ }\textbf {\bibinfo {volume} {66}},\ \bibinfo {pages} {2072} (\bibinfo {year} {2021})}\BibitemShut {NoStop}%
\bibitem [{\citenamefont {Chai}\ \emph {et~al.}(2022)\citenamefont {Chai}, \citenamefont {Chen}, \citenamefont {Zha}, \citenamefont {Pei},\ and\ \citenamefont {Xu}}]{peisy21}%
  \BibitemOpen
  \bibfield  {author} {\bibinfo {author} {\bibfnamefont {Q.}~\bibnamefont {Chai}}, \bibinfo {author} {\bibfnamefont {H.}~\bibnamefont {Chen}}, \bibinfo {author} {\bibfnamefont {M.}~\bibnamefont {Zha}}, \bibinfo {author} {\bibfnamefont {J.}~\bibnamefont {Pei}},\ and\ \bibinfo {author} {\bibfnamefont {F.}~\bibnamefont {Xu}},\ }\bibfield  {journal} {\bibinfo  {journal} {Symmetry}\ }\textbf {\bibinfo {volume} {14}},\ \href {https://doi.org/10.3390/sym14020215} {10.3390/sym14020215} (\bibinfo {year} {2022})\BibitemShut {NoStop}%
\bibitem [{\citenamefont {Zhang}\ \emph {et~al.}(2023{\natexlab{a}})\citenamefont {Zhang}, \citenamefont {Papakonstantinou}, \citenamefont {Mun}, \citenamefont {Kim}, \citenamefont {Yan},\ and\ \citenamefont {Sun}}]{ZhangKY23}%
  \BibitemOpen
  \bibfield  {author} {\bibinfo {author} {\bibfnamefont {K.~Y.}\ \bibnamefont {Zhang}}, \bibinfo {author} {\bibfnamefont {P.}~\bibnamefont {Papakonstantinou}}, \bibinfo {author} {\bibfnamefont {M.-H.}\ \bibnamefont {Mun}}, \bibinfo {author} {\bibfnamefont {Y.}~\bibnamefont {Kim}}, \bibinfo {author} {\bibfnamefont {H.}~\bibnamefont {Yan}},\ and\ \bibinfo {author} {\bibfnamefont {X.-X.}\ \bibnamefont {Sun}},\ }\href {https://doi.org/10.1103/PhysRevC.107.L041303} {\bibfield  {journal} {\bibinfo  {journal} {Phys. Rev. C}\ }\textbf {\bibinfo {volume} {107}},\ \bibinfo {pages} {L041303} (\bibinfo {year} {2023}{\natexlab{a}})}\BibitemShut {NoStop}%
\bibitem [{\citenamefont {Zhang}\ \emph {et~al.}(2023{\natexlab{b}})\citenamefont {Zhang}, \citenamefont {Yang}, \citenamefont {An}, \citenamefont {Zhang}, \citenamefont {Papakonstantinou}, \citenamefont {Mun}, \citenamefont {Kim},\ and\ \citenamefont {Yan}}]{ZHANG2023138112}%
  \BibitemOpen
  \bibfield  {author} {\bibinfo {author} {\bibfnamefont {K.}~\bibnamefont {Zhang}}, \bibinfo {author} {\bibfnamefont {S.}~\bibnamefont {Yang}}, \bibinfo {author} {\bibfnamefont {J.}~\bibnamefont {An}}, \bibinfo {author} {\bibfnamefont {S.}~\bibnamefont {Zhang}}, \bibinfo {author} {\bibfnamefont {P.}~\bibnamefont {Papakonstantinou}}, \bibinfo {author} {\bibfnamefont {M.-H.}\ \bibnamefont {Mun}}, \bibinfo {author} {\bibfnamefont {Y.}~\bibnamefont {Kim}},\ and\ \bibinfo {author} {\bibfnamefont {H.}~\bibnamefont {Yan}},\ }\href {https://doi.org/https://doi.org/10.1016/j.physletb.2023.138112} {\bibfield  {journal} {\bibinfo  {journal} {Phys. Lett. B}\ }\textbf {\bibinfo {volume} {844}},\ \bibinfo {pages} {138112} (\bibinfo {year} {2023}{\natexlab{b}})}\BibitemShut {NoStop}%
\bibitem [{\citenamefont {An}\ \emph {et~al.}(2024)\citenamefont {An}, \citenamefont {Zhang}, \citenamefont {Lu}, \citenamefont {Zhong},\ and\ \citenamefont {Zhang}}]{plbzhang2024}%
  \BibitemOpen
  \bibfield  {author} {\bibinfo {author} {\bibfnamefont {J.-L.}\ \bibnamefont {An}}, \bibinfo {author} {\bibfnamefont {K.-Y.}\ \bibnamefont {Zhang}}, \bibinfo {author} {\bibfnamefont {Q.}~\bibnamefont {Lu}}, \bibinfo {author} {\bibfnamefont {S.-Y.}\ \bibnamefont {Zhong}},\ and\ \bibinfo {author} {\bibfnamefont {S.-S.}\ \bibnamefont {Zhang}},\ }\href {https://doi.org/https://doi.org/10.1016/j.physletb.2023.138422} {\bibfield  {journal} {\bibinfo  {journal} {Phys. Lett. B}\ }\textbf {\bibinfo {volume} {849}},\ \bibinfo {pages} {138422} (\bibinfo {year} {2024})}\BibitemShut {NoStop}%
\bibitem [{\citenamefont {Takatsu}\ \emph {et~al.}(2023)\citenamefont {Takatsu}, \citenamefont {Suzuki}, \citenamefont {Horiuchi},\ and\ \citenamefont {Kimura}}]{Takatsu2023}%
  \BibitemOpen
  \bibfield  {author} {\bibinfo {author} {\bibfnamefont {R.}~\bibnamefont {Takatsu}}, \bibinfo {author} {\bibfnamefont {Y.}~\bibnamefont {Suzuki}}, \bibinfo {author} {\bibfnamefont {W.}~\bibnamefont {Horiuchi}},\ and\ \bibinfo {author} {\bibfnamefont {M.}~\bibnamefont {Kimura}},\ }\href {https://doi.org/10.1103/PhysRevC.107.024314} {\bibfield  {journal} {\bibinfo  {journal} {Phys. Rev. C}\ }\textbf {\bibinfo {volume} {107}},\ \bibinfo {pages} {024314} (\bibinfo {year} {2023})}\BibitemShut {NoStop}%
\bibitem [{\citenamefont {Hamamoto}(2019)}]{Hamamoto2019}%
  \BibitemOpen
  \bibfield  {author} {\bibinfo {author} {\bibfnamefont {I.}~\bibnamefont {Hamamoto}},\ }\href {https://doi.org/10.1103/PhysRevC.100.014324} {\bibfield  {journal} {\bibinfo  {journal} {Phys. Rev. C}\ }\textbf {\bibinfo {volume} {100}},\ \bibinfo {pages} {014324} (\bibinfo {year} {2019})}\BibitemShut {NoStop}%
\bibitem [{\citenamefont {Yanagisawa}\ \emph {et~al.}(2003)\citenamefont {Yanagisawa}, \citenamefont {Notani}, \citenamefont {Sakurai}, \citenamefont {Kunibu}, \citenamefont {Akiyoshi}, \citenamefont {Aoi}, \citenamefont {Baba}, \citenamefont {Demichi}, \citenamefont {Fukuda}, \citenamefont {Hasegawa}, \citenamefont {Higurashi}, \citenamefont {Ishihara}, \citenamefont {Iwasa}, \citenamefont {Iwasaki}, \citenamefont {Gomi}, \citenamefont {Kanno}, \citenamefont {Kurokawa}, \citenamefont {Matsuyama}, \citenamefont {Michimasa}, \citenamefont {Minemura}, \citenamefont {Mizoi}, \citenamefont {Nakamura}, \citenamefont {Saito}, \citenamefont {Serata}, \citenamefont {Shimoura}, \citenamefont {Sugimoto}, \citenamefont {Takeshita}, \citenamefont {Takeuchi}, \citenamefont {Ue}, \citenamefont {Yamada}, \citenamefont {Yoneda},\ and\ \citenamefont {Motobayashi}}]{Yanagisawa2003}%
  \BibitemOpen
  \bibfield  {author} {\bibinfo {author} {\bibfnamefont {Y.}~\bibnamefont {Yanagisawa}}, \bibinfo {author} {\bibfnamefont {M.}~\bibnamefont {Notani}}, \bibinfo {author} {\bibfnamefont {H.}~\bibnamefont {Sakurai}}, \bibinfo {author} {\bibfnamefont {M.}~\bibnamefont {Kunibu}}, \bibinfo {author} {\bibfnamefont {H.}~\bibnamefont {Akiyoshi}}, \bibinfo {author} {\bibfnamefont {N.}~\bibnamefont {Aoi}}, \bibinfo {author} {\bibfnamefont {H.}~\bibnamefont {Baba}}, \bibinfo {author} {\bibfnamefont {K.}~\bibnamefont {Demichi}}, \bibinfo {author} {\bibfnamefont {N.}~\bibnamefont {Fukuda}}, \bibinfo {author} {\bibfnamefont {H.}~\bibnamefont {Hasegawa}}, \bibinfo {author} {\bibfnamefont {Y.}~\bibnamefont {Higurashi}}, \bibinfo {author} {\bibfnamefont {M.}~\bibnamefont {Ishihara}}, \bibinfo {author} {\bibfnamefont {N.}~\bibnamefont {Iwasa}}, \bibinfo {author} {\bibfnamefont {H.}~\bibnamefont {Iwasaki}}, \bibinfo {author} {\bibfnamefont {T.}~\bibnamefont {Gomi}}, \bibinfo {author} {\bibfnamefont {S.}~\bibnamefont {Kanno}},
  \bibinfo {author} {\bibfnamefont {M.}~\bibnamefont {Kurokawa}}, \bibinfo {author} {\bibfnamefont {Y.}~\bibnamefont {Matsuyama}}, \bibinfo {author} {\bibfnamefont {S.}~\bibnamefont {Michimasa}}, \bibinfo {author} {\bibfnamefont {T.}~\bibnamefont {Minemura}}, \bibinfo {author} {\bibfnamefont {T.}~\bibnamefont {Mizoi}}, \bibinfo {author} {\bibfnamefont {T.}~\bibnamefont {Nakamura}}, \bibinfo {author} {\bibfnamefont {A.}~\bibnamefont {Saito}}, \bibinfo {author} {\bibfnamefont {M.}~\bibnamefont {Serata}}, \bibinfo {author} {\bibfnamefont {S.}~\bibnamefont {Shimoura}}, \bibinfo {author} {\bibfnamefont {T.}~\bibnamefont {Sugimoto}}, \bibinfo {author} {\bibfnamefont {E.}~\bibnamefont {Takeshita}}, \bibinfo {author} {\bibfnamefont {S.}~\bibnamefont {Takeuchi}}, \bibinfo {author} {\bibfnamefont {K.}~\bibnamefont {Ue}}, \bibinfo {author} {\bibfnamefont {K.}~\bibnamefont {Yamada}}, \bibinfo {author} {\bibfnamefont {K.}~\bibnamefont {Yoneda}},\ and\ \bibinfo {author} {\bibfnamefont {T.}~\bibnamefont {Motobayashi}},\
  }\href {https://doi.org/https://doi.org/10.1016/S0370-2693(03)00802-5} {\bibfield  {journal} {\bibinfo  {journal} {Phys. Lett. B}\ }\textbf {\bibinfo {volume} {566}},\ \bibinfo {pages} {84} (\bibinfo {year} {2003})}\BibitemShut {NoStop}%
\bibitem [{\citenamefont {Doornenbal}\ \emph {et~al.}(2009)\citenamefont {Doornenbal}, \citenamefont {Scheit}, \citenamefont {Aoi}, \citenamefont {Takeuchi}, \citenamefont {Li}, \citenamefont {Takeshita}, \citenamefont {Wang}, \citenamefont {Baba}, \citenamefont {Deguchi}, \citenamefont {Fukuda}, \citenamefont {Geissel}, \citenamefont {Gernh\"auser}, \citenamefont {Gibelin}, \citenamefont {Hachiuma}, \citenamefont {Hara}, \citenamefont {Hinke}, \citenamefont {Inabe}, \citenamefont {Itahashi}, \citenamefont {Itoh}, \citenamefont {Kameda}, \citenamefont {Kanno}, \citenamefont {Kawada}, \citenamefont {Kobayashi}, \citenamefont {Kondo}, \citenamefont {Kr\"ucken}, \citenamefont {Kubo}, \citenamefont {Kuboki}, \citenamefont {Kusaka}, \citenamefont {Lantz}, \citenamefont {Michimasa}, \citenamefont {Motobayashi}, \citenamefont {Nakamura}, \citenamefont {Nakao}, \citenamefont {Namihira}, \citenamefont {Nishimura}, \citenamefont {Ohnishi}, \citenamefont {Ohtake}, \citenamefont {Orr}, \citenamefont {Otsu}, \citenamefont
  {Ozeki}, \citenamefont {Satou}, \citenamefont {Shimoura}, \citenamefont {Sumikama}, \citenamefont {Takechi}, \citenamefont {Takeda}, \citenamefont {Tanaka}, \citenamefont {Tanaka}, \citenamefont {Togano}, \citenamefont {Winkler}, \citenamefont {Yanagisawa}, \citenamefont {Yoneda}, \citenamefont {Yoshida}, \citenamefont {Yoshida},\ and\ \citenamefont {Sakurai}}]{Door2009}%
  \BibitemOpen
  \bibfield  {author} {\bibinfo {author} {\bibfnamefont {P.}~\bibnamefont {Doornenbal}}, \bibinfo {author} {\bibfnamefont {H.}~\bibnamefont {Scheit}}, \bibinfo {author} {\bibfnamefont {N.}~\bibnamefont {Aoi}}, \bibinfo {author} {\bibfnamefont {S.}~\bibnamefont {Takeuchi}}, \bibinfo {author} {\bibfnamefont {K.}~\bibnamefont {Li}}, \bibinfo {author} {\bibfnamefont {E.}~\bibnamefont {Takeshita}}, \bibinfo {author} {\bibfnamefont {H.}~\bibnamefont {Wang}}, \bibinfo {author} {\bibfnamefont {H.}~\bibnamefont {Baba}}, \bibinfo {author} {\bibfnamefont {S.}~\bibnamefont {Deguchi}}, \bibinfo {author} {\bibfnamefont {N.}~\bibnamefont {Fukuda}}, \bibinfo {author} {\bibfnamefont {H.}~\bibnamefont {Geissel}}, \bibinfo {author} {\bibfnamefont {R.}~\bibnamefont {Gernh\"auser}}, \bibinfo {author} {\bibfnamefont {J.}~\bibnamefont {Gibelin}}, \bibinfo {author} {\bibfnamefont {I.}~\bibnamefont {Hachiuma}}, \bibinfo {author} {\bibfnamefont {Y.}~\bibnamefont {Hara}}, \bibinfo {author} {\bibfnamefont {C.}~\bibnamefont {Hinke}},
  \bibinfo {author} {\bibfnamefont {N.}~\bibnamefont {Inabe}}, \bibinfo {author} {\bibfnamefont {K.}~\bibnamefont {Itahashi}}, \bibinfo {author} {\bibfnamefont {S.}~\bibnamefont {Itoh}}, \bibinfo {author} {\bibfnamefont {D.}~\bibnamefont {Kameda}}, \bibinfo {author} {\bibfnamefont {S.}~\bibnamefont {Kanno}}, \bibinfo {author} {\bibfnamefont {Y.}~\bibnamefont {Kawada}}, \bibinfo {author} {\bibfnamefont {N.}~\bibnamefont {Kobayashi}}, \bibinfo {author} {\bibfnamefont {Y.}~\bibnamefont {Kondo}}, \bibinfo {author} {\bibfnamefont {R.}~\bibnamefont {Kr\"ucken}}, \bibinfo {author} {\bibfnamefont {T.}~\bibnamefont {Kubo}}, \bibinfo {author} {\bibfnamefont {T.}~\bibnamefont {Kuboki}}, \bibinfo {author} {\bibfnamefont {K.}~\bibnamefont {Kusaka}}, \bibinfo {author} {\bibfnamefont {M.}~\bibnamefont {Lantz}}, \bibinfo {author} {\bibfnamefont {S.}~\bibnamefont {Michimasa}}, \bibinfo {author} {\bibfnamefont {T.}~\bibnamefont {Motobayashi}}, \bibinfo {author} {\bibfnamefont {T.}~\bibnamefont {Nakamura}}, \bibinfo {author}
  {\bibfnamefont {T.}~\bibnamefont {Nakao}}, \bibinfo {author} {\bibfnamefont {K.}~\bibnamefont {Namihira}}, \bibinfo {author} {\bibfnamefont {S.}~\bibnamefont {Nishimura}}, \bibinfo {author} {\bibfnamefont {T.}~\bibnamefont {Ohnishi}}, \bibinfo {author} {\bibfnamefont {M.}~\bibnamefont {Ohtake}}, \bibinfo {author} {\bibfnamefont {N.~A.}\ \bibnamefont {Orr}}, \bibinfo {author} {\bibfnamefont {H.}~\bibnamefont {Otsu}}, \bibinfo {author} {\bibfnamefont {K.}~\bibnamefont {Ozeki}}, \bibinfo {author} {\bibfnamefont {Y.}~\bibnamefont {Satou}}, \bibinfo {author} {\bibfnamefont {S.}~\bibnamefont {Shimoura}}, \bibinfo {author} {\bibfnamefont {T.}~\bibnamefont {Sumikama}}, \bibinfo {author} {\bibfnamefont {M.}~\bibnamefont {Takechi}}, \bibinfo {author} {\bibfnamefont {H.}~\bibnamefont {Takeda}}, \bibinfo {author} {\bibfnamefont {K.~N.}\ \bibnamefont {Tanaka}}, \bibinfo {author} {\bibfnamefont {K.}~\bibnamefont {Tanaka}}, \bibinfo {author} {\bibfnamefont {Y.}~\bibnamefont {Togano}}, \bibinfo {author} {\bibfnamefont
  {M.}~\bibnamefont {Winkler}}, \bibinfo {author} {\bibfnamefont {Y.}~\bibnamefont {Yanagisawa}}, \bibinfo {author} {\bibfnamefont {K.}~\bibnamefont {Yoneda}}, \bibinfo {author} {\bibfnamefont {A.}~\bibnamefont {Yoshida}}, \bibinfo {author} {\bibfnamefont {K.}~\bibnamefont {Yoshida}},\ and\ \bibinfo {author} {\bibfnamefont {H.}~\bibnamefont {Sakurai}},\ }\href {https://doi.org/10.1103/PhysRevLett.103.032501} {\bibfield  {journal} {\bibinfo  {journal} {Phys. Rev. Lett.}\ }\textbf {\bibinfo {volume} {103}},\ \bibinfo {pages} {032501} (\bibinfo {year} {2009})}\BibitemShut {NoStop}%
\bibitem [{\citenamefont {Yordanov}\ \emph {et~al.}(2007)\citenamefont {Yordanov}, \citenamefont {Kowalska}, \citenamefont {Blaum}, \citenamefont {De~Rydt}, \citenamefont {Flanagan}, \citenamefont {Lievens}, \citenamefont {Neugart}, \citenamefont {Neyens},\ and\ \citenamefont {Stroke}}]{Yordanov2007}%
  \BibitemOpen
  \bibfield  {author} {\bibinfo {author} {\bibfnamefont {D.~T.}\ \bibnamefont {Yordanov}}, \bibinfo {author} {\bibfnamefont {M.}~\bibnamefont {Kowalska}}, \bibinfo {author} {\bibfnamefont {K.}~\bibnamefont {Blaum}}, \bibinfo {author} {\bibfnamefont {M.}~\bibnamefont {De~Rydt}}, \bibinfo {author} {\bibfnamefont {K.~T.}\ \bibnamefont {Flanagan}}, \bibinfo {author} {\bibfnamefont {P.}~\bibnamefont {Lievens}}, \bibinfo {author} {\bibfnamefont {R.}~\bibnamefont {Neugart}}, \bibinfo {author} {\bibfnamefont {G.}~\bibnamefont {Neyens}},\ and\ \bibinfo {author} {\bibfnamefont {H.~H.}\ \bibnamefont {Stroke}},\ }\href {https://doi.org/10.1103/PhysRevLett.99.212501} {\bibfield  {journal} {\bibinfo  {journal} {Phys. Rev. Lett.}\ }\textbf {\bibinfo {volume} {99}},\ \bibinfo {pages} {212501} (\bibinfo {year} {2007})}\BibitemShut {NoStop}%
\bibitem [{\citenamefont {Zhou}\ \emph {et~al.}(2003)\citenamefont {Zhou}, \citenamefont {Meng},\ and\ \citenamefont {Ring}}]{zhou03}%
  \BibitemOpen
  \bibfield  {author} {\bibinfo {author} {\bibfnamefont {S.-G.}\ \bibnamefont {Zhou}}, \bibinfo {author} {\bibfnamefont {J.}~\bibnamefont {Meng}},\ and\ \bibinfo {author} {\bibfnamefont {P.}~\bibnamefont {Ring}},\ }\href {https://doi.org/10.1103/PhysRevC.68.034323} {\bibfield  {journal} {\bibinfo  {journal} {Phys. Rev. C}\ }\textbf {\bibinfo {volume} {68}},\ \bibinfo {pages} {034323} (\bibinfo {year} {2003})}\BibitemShut {NoStop}%
\bibitem [{\citenamefont {Poves}\ and\ \citenamefont {Retamosa}(1994)}]{POVES1994}%
  \BibitemOpen
  \bibfield  {author} {\bibinfo {author} {\bibfnamefont {A.}~\bibnamefont {Poves}}\ and\ \bibinfo {author} {\bibfnamefont {J.}~\bibnamefont {Retamosa}},\ }\href {https://doi.org/https://doi.org/10.1016/0375-9474(94)90058-2} {\bibfield  {journal} {\bibinfo  {journal} {Nucl. Phys. A}\ }\textbf {\bibinfo {volume} {571}},\ \bibinfo {pages} {221} (\bibinfo {year} {1994})}\BibitemShut {NoStop}%
\bibitem [{\citenamefont {Doornenbal}\ \emph {et~al.}(2016)\citenamefont {Doornenbal}, \citenamefont {Scheit}, \citenamefont {Takeuchi}, \citenamefont {Aoi}, \citenamefont {Li}, \citenamefont {Matsushita}, \citenamefont {Steppenbeck}, \citenamefont {Wang}, \citenamefont {Baba}, \citenamefont {Ideguchi}, \citenamefont {Kobayashi}, \citenamefont {Kondo}, \citenamefont {Lee}, \citenamefont {Michimasa}, \citenamefont {Motobayashi}, \citenamefont {Poves}, \citenamefont {Sakurai}, \citenamefont {Takechi}, \citenamefont {Togano},\ and\ \citenamefont {Yoneda}}]{mg36-2}%
  \BibitemOpen
  \bibfield  {author} {\bibinfo {author} {\bibfnamefont {P.}~\bibnamefont {Doornenbal}}, \bibinfo {author} {\bibfnamefont {H.}~\bibnamefont {Scheit}}, \bibinfo {author} {\bibfnamefont {S.}~\bibnamefont {Takeuchi}}, \bibinfo {author} {\bibfnamefont {N.}~\bibnamefont {Aoi}}, \bibinfo {author} {\bibfnamefont {K.}~\bibnamefont {Li}}, \bibinfo {author} {\bibfnamefont {M.}~\bibnamefont {Matsushita}}, \bibinfo {author} {\bibfnamefont {D.}~\bibnamefont {Steppenbeck}}, \bibinfo {author} {\bibfnamefont {H.}~\bibnamefont {Wang}}, \bibinfo {author} {\bibfnamefont {H.}~\bibnamefont {Baba}}, \bibinfo {author} {\bibfnamefont {E.}~\bibnamefont {Ideguchi}}, \bibinfo {author} {\bibfnamefont {N.}~\bibnamefont {Kobayashi}}, \bibinfo {author} {\bibfnamefont {Y.}~\bibnamefont {Kondo}}, \bibinfo {author} {\bibfnamefont {J.}~\bibnamefont {Lee}}, \bibinfo {author} {\bibfnamefont {S.}~\bibnamefont {Michimasa}}, \bibinfo {author} {\bibfnamefont {T.}~\bibnamefont {Motobayashi}}, \bibinfo {author} {\bibfnamefont {A.}~\bibnamefont
  {Poves}}, \bibinfo {author} {\bibfnamefont {H.}~\bibnamefont {Sakurai}}, \bibinfo {author} {\bibfnamefont {M.}~\bibnamefont {Takechi}}, \bibinfo {author} {\bibfnamefont {Y.}~\bibnamefont {Togano}},\ and\ \bibinfo {author} {\bibfnamefont {K.}~\bibnamefont {Yoneda}},\ }\href {https://doi.org/10.1103/PhysRevC.93.044306} {\bibfield  {journal} {\bibinfo  {journal} {Phys. Rev. C}\ }\textbf {\bibinfo {volume} {93}},\ \bibinfo {pages} {044306} (\bibinfo {year} {2016})}\BibitemShut {NoStop}%
\bibitem [{\citenamefont {Doornenbal}\ \emph {et~al.}(2013)\citenamefont {Doornenbal}, \citenamefont {Scheit}, \citenamefont {Takeuchi}, \citenamefont {Aoi}, \citenamefont {Li}, \citenamefont {Matsushita}, \citenamefont {Steppenbeck}, \citenamefont {Wang}, \citenamefont {Baba}, \citenamefont {Crawford}, \citenamefont {Hoffman}, \citenamefont {Hughes}, \citenamefont {Ideguchi}, \citenamefont {Kobayashi}, \citenamefont {Kondo}, \citenamefont {Lee}, \citenamefont {Michimasa}, \citenamefont {Motobayashi}, \citenamefont {Sakurai}, \citenamefont {Takechi}, \citenamefont {Togano}, \citenamefont {Winkler},\ and\ \citenamefont {Yoneda}}]{mg36-mg38}%
  \BibitemOpen
  \bibfield  {author} {\bibinfo {author} {\bibfnamefont {P.}~\bibnamefont {Doornenbal}}, \bibinfo {author} {\bibfnamefont {H.}~\bibnamefont {Scheit}}, \bibinfo {author} {\bibfnamefont {S.}~\bibnamefont {Takeuchi}}, \bibinfo {author} {\bibfnamefont {N.}~\bibnamefont {Aoi}}, \bibinfo {author} {\bibfnamefont {K.}~\bibnamefont {Li}}, \bibinfo {author} {\bibfnamefont {M.}~\bibnamefont {Matsushita}}, \bibinfo {author} {\bibfnamefont {D.}~\bibnamefont {Steppenbeck}}, \bibinfo {author} {\bibfnamefont {H.}~\bibnamefont {Wang}}, \bibinfo {author} {\bibfnamefont {H.}~\bibnamefont {Baba}}, \bibinfo {author} {\bibfnamefont {H.}~\bibnamefont {Crawford}}, \bibinfo {author} {\bibfnamefont {C.~R.}\ \bibnamefont {Hoffman}}, \bibinfo {author} {\bibfnamefont {R.}~\bibnamefont {Hughes}}, \bibinfo {author} {\bibfnamefont {E.}~\bibnamefont {Ideguchi}}, \bibinfo {author} {\bibfnamefont {N.}~\bibnamefont {Kobayashi}}, \bibinfo {author} {\bibfnamefont {Y.}~\bibnamefont {Kondo}}, \bibinfo {author} {\bibfnamefont {J.}~\bibnamefont
  {Lee}}, \bibinfo {author} {\bibfnamefont {S.}~\bibnamefont {Michimasa}}, \bibinfo {author} {\bibfnamefont {T.}~\bibnamefont {Motobayashi}}, \bibinfo {author} {\bibfnamefont {H.}~\bibnamefont {Sakurai}}, \bibinfo {author} {\bibfnamefont {M.}~\bibnamefont {Takechi}}, \bibinfo {author} {\bibfnamefont {Y.}~\bibnamefont {Togano}}, \bibinfo {author} {\bibfnamefont {R.}~\bibnamefont {Winkler}},\ and\ \bibinfo {author} {\bibfnamefont {K.}~\bibnamefont {Yoneda}},\ }\href {https://doi.org/10.1103/PhysRevLett.111.212502} {\bibfield  {journal} {\bibinfo  {journal} {Phys. Rev. Lett.}\ }\textbf {\bibinfo {volume} {111}},\ \bibinfo {pages} {212502} (\bibinfo {year} {2013})}\BibitemShut {NoStop}%
\bibitem [{\citenamefont {Ren}\ \emph {et~al.}(1996)\citenamefont {Ren}, \citenamefont {Zhu}, \citenamefont {Cai},\ and\ \citenamefont {Xu}}]{REN1996241}%
  \BibitemOpen
  \bibfield  {author} {\bibinfo {author} {\bibfnamefont {Z.}~\bibnamefont {Ren}}, \bibinfo {author} {\bibfnamefont {Z.}~\bibnamefont {Zhu}}, \bibinfo {author} {\bibfnamefont {Y.}~\bibnamefont {Cai}},\ and\ \bibinfo {author} {\bibfnamefont {G.}~\bibnamefont {Xu}},\ }\href {https://doi.org/https://doi.org/10.1016/0370-2693(96)00462-5} {\bibfield  {journal} {\bibinfo  {journal} {Phys. Lett. B}\ }\textbf {\bibinfo {volume} {380}},\ \bibinfo {pages} {241} (\bibinfo {year} {1996})}\BibitemShut {NoStop}%
\bibitem [{\citenamefont {Jin-Gen}\ \emph {et~al.}(2005)\citenamefont {Jin-Gen}, \citenamefont {Xiang-Zhou}, \citenamefont {Ting-Tai}, \citenamefont {Yu-Gang}, \citenamefont {Zhong-Zhou}, \citenamefont {De-Qing}, \citenamefont {Chen}, \citenamefont {Yi-Bin}, \citenamefont {Wei}, \citenamefont {Xing-Fei}, \citenamefont {Kun}, \citenamefont {Guo-Liang}, \citenamefont {Wen-Dong}, \citenamefont {Jin-Hui}, \citenamefont {Ting-Zhi}, \citenamefont {Jia-Xu}, \citenamefont {Chun-Wang},\ and\ \citenamefont {Wen-Qing}}]{Chen2005}%
  \BibitemOpen
  \bibfield  {author} {\bibinfo {author} {\bibfnamefont {C.}~\bibnamefont {Jin-Gen}}, \bibinfo {author} {\bibfnamefont {C.}~\bibnamefont {Xiang-Zhou}}, \bibinfo {author} {\bibfnamefont {W.}~\bibnamefont {Ting-Tai}}, \bibinfo {author} {\bibfnamefont {M.}~\bibnamefont {Yu-Gang}}, \bibinfo {author} {\bibfnamefont {R.}~\bibnamefont {Zhong-Zhou}}, \bibinfo {author} {\bibfnamefont {F.}~\bibnamefont {De-Qing}}, \bibinfo {author} {\bibfnamefont {Z.}~\bibnamefont {Chen}}, \bibinfo {author} {\bibfnamefont {W.}~\bibnamefont {Yi-Bin}}, \bibinfo {author} {\bibfnamefont {G.}~\bibnamefont {Wei}}, \bibinfo {author} {\bibfnamefont {Z.}~\bibnamefont {Xing-Fei}}, \bibinfo {author} {\bibfnamefont {W.}~\bibnamefont {Kun}}, \bibinfo {author} {\bibfnamefont {M.}~\bibnamefont {Guo-Liang}}, \bibinfo {author} {\bibfnamefont {T.}~\bibnamefont {Wen-Dong}}, \bibinfo {author} {\bibfnamefont {C.}~\bibnamefont {Jin-Hui}}, \bibinfo {author} {\bibfnamefont {Y.}~\bibnamefont {Ting-Zhi}}, \bibinfo {author} {\bibfnamefont {Z.}~\bibnamefont
  {Jia-Xu}}, \bibinfo {author} {\bibfnamefont {M.}~\bibnamefont {Chun-Wang}},\ and\ \bibinfo {author} {\bibfnamefont {S.}~\bibnamefont {Wen-Qing}},\ }\href {https://doi.org/10.1088/1009-1963/14/12/013} {\bibfield  {journal} {\bibinfo  {journal} {Chin. Phys.}\ }\textbf {\bibinfo {volume} {14}},\ \bibinfo {pages} {2444} (\bibinfo {year} {2005})}\BibitemShut {NoStop}%
\bibitem [{\citenamefont {Xiong}\ \emph {et~al.}(2016)\citenamefont {Xiong}, \citenamefont {Pei}, \citenamefont {Zhang},\ and\ \citenamefont {Zhu}}]{Xiong_2016}%
  \BibitemOpen
  \bibfield  {author} {\bibinfo {author} {\bibfnamefont {X.-Y.}\ \bibnamefont {Xiong}}, \bibinfo {author} {\bibfnamefont {J.-C.}\ \bibnamefont {Pei}}, \bibinfo {author} {\bibfnamefont {Y.-N.}\ \bibnamefont {Zhang}},\ and\ \bibinfo {author} {\bibfnamefont {Y.}~\bibnamefont {Zhu}},\ }\href {https://doi.org/10.1088/1674-1137/40/2/024101} {\bibfield  {journal} {\bibinfo  {journal} {Chin. Phys. C}\ }\textbf {\bibinfo {volume} {40}},\ \bibinfo {pages} {024101} (\bibinfo {year} {2016})}\BibitemShut {NoStop}%
\bibitem [{\citenamefont {Sun}\ and\ \citenamefont {Meng}(2022)}]{PhysRevC.105.044312}%
  \BibitemOpen
  \bibfield  {author} {\bibinfo {author} {\bibfnamefont {X.}~\bibnamefont {Sun}}\ and\ \bibinfo {author} {\bibfnamefont {J.}~\bibnamefont {Meng}},\ }\href {https://doi.org/10.1103/PhysRevC.105.044312} {\bibfield  {journal} {\bibinfo  {journal} {Phys. Rev. C}\ }\textbf {\bibinfo {volume} {105}},\ \bibinfo {pages} {044312} (\bibinfo {year} {2022})}\BibitemShut {NoStop}%
\end{thebibliography}%
\end{document}